%% file: pwaml_arXivX.tex
\documentclass[letter,preprint,aps,prl,nofootinbib,longbibliography]{revtex4-1}
\usepackage{enumitem}
\usepackage{bm}
\usepackage[colorlinks = true,citecolor=red]{hyperref}
\usepackage{amsmath}
\usepackage{graphicx}
\usepackage{epsfig}

\input{defs}

\input{defs2}

\newcommand{\bS}{{\bf S}}
\begin{document}

\title{Phil Anderson's Magnetic Ideas in Science}\label{ra_ch1}

\author{Piers Coleman$^{1,2}$}

\address{$^1$ Center for Materials Theory, Department of 
Physics and Astronomy, Rutgers University, 
Piscataway, NJ 08854}
\address{$^{2}$
Department of Physics, Royal Holloway, University of London, Egham, Surrey TW20 0EX, UK}

\begin{abstract}
In Philip W. Anderson's research, 
magnetism has always played a special role, 
providing a prism through which 
other more complex forms of collective behavior and broken symmetry
could be examined. I discuss his 
work on magnetism from the 1950s,
where his
early work on antiferromagnetism led to the pseudospin treatment of
superconductivity - to the 70s and 80s, 
highlighting 
his contribution to the physics of local magnetic moments. 
Phil's interest in the mechanism of moment formation, and screening
evolved into the modern theory of the Kondo effect and heavy fermions.
\\
\\

\noindent Reprinted with permission from {\sl ``PWA90: A Lifetime of
Emergence''}, editors P. Chandra, P. Coleman, G. Kotliar, P. Ong,
D. Stein and C. Yu, pp 187-213, World Scientific (2016).
\end{abstract}

\maketitle

\section{Introduction}

This article is based on a talk I gave about Phil Anderson's
contributions to our understanding of magnetism and its links with
superconductivity, at the 110th Rutgers Statistical mechanics
meeting. This event, organized by Joel Lebowitz, was a continuation of
the New Jersey celebrations began at ``PWA90: A lifetime in
emergence'', on the weekend of Phil Anderson's 90th birthday in
December 2013.  My title has a double-entendre, for Phil's ideas in
science have a magnetic quality, and have long provided inspiration,
attracting students such as myself, to work with him. I first learned
about Phil Anderson as an undergraduate at Cambridge in 1979, some
three years after he had left for Princeton. Phil had left behind many
legends at Cambridge, one of which was that he had ideas of 
depth and great beauty, but also that he was very hard to understand. 
For me, as with many fellow
students of Phil, the thought of working with an advisor with 
some of the best ideas on the block 
was very attractive, and it was {\sl this} magnetism that brought me
over to New York Harbor nine months later, to start a Ph. D. with Phil
at Princeton.  

One of the recurrent themes of Anderson's work, is the
importance of using models as a gateway to discovering general mechanisms
and principles, and throughout his career, models of magnetism played
a key role. 
In his book ``Basic concepts of condensed
matter physics''\cite{andersonconcepts}, 
Anderson gives various examples of such basic principles, 
such as adiabatic continuation, 
the idea of renormalization 
as a way to eliminate all but the
essential degrees of freedom, and most famously, the 
link between 
broken symmetry and the idea of
{\sl generalized rigidity}, writing
\begin{quotation}
{\sl ``We are so accustomed to the rigidity of solid bodies— that is hard to realize that such action at a distance is not built into the laws of nature … It is strictly a consequence of the fact that the energy is minimized when symmetry is broken in the same way throughout the sample… 

The generalization of this concept to all instances of broken symmetry is what I call generalized rigidity. It is responsible for most of the unique properties of the ordered (broken-symmetry) states: ferromagnetism, superconductivity, superfluidity, etc.''}
\end{quotation}

Yet in the 50s, when Phil began working on magnetism, these 
ideas had not yet been formed: the term 
{\sl broken symmetry} was not yet in common usage, 
renormalization was little more than a method of
eliminating divergences in particle physics and 
beyond the Ising and Heisenberg models, there were
almost no other simple models for interacting electrons.  Phil's
studies of models of magnetism spanning the next three decades played 
a central role in the development of his thoughts on general principles
and mechanisms in condensed matter physics, especially those
underlying broken symmetry. 

I'll  discuss  three main periods in Phil's work as 
shown in the time-line of Fig. 1, and arbitrarily color coded as the
``blue'', ``orange'' and ``green'' period. 
My short presentation is unfortunately
highly selective but I hope it will give a
useful flavor to the reader 
of the evolution of ideas that have accompanied Phil's
work in magnetism.
\fight=\textwidth
\fg{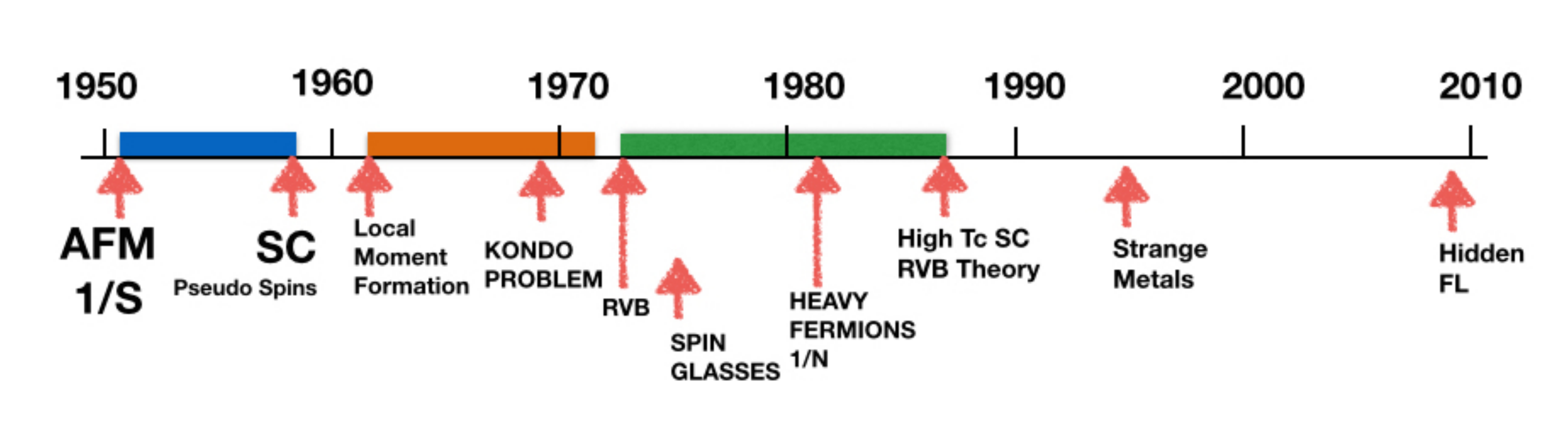}{fig1}{Three periods of Anderson's
research into magnetism  selectively discussed in this article. 
Blue period: from antiferromagnetism to superconductivity. Orange
period: theory of local moment formation and the Kondo problem. Green
period: from resonating valence
bonds (RVB) to high temperature superconductivity. 
}

\section{Blue Period: Antiferromagnetism and Superconductivity}\label{}

Today it is hard to imagine the uncertainties connected with 
antiferromagnetism and broken symmetry 
around 1950.  While N\'eel and Landau\cite{Neel,Landau} had 
independently predicted  ``antiferromagnetism'', with a staggered
magnetization ($
\uparrow\downarrow 
\uparrow\downarrow 
\uparrow\downarrow 
$), as the classical ground-state of the 
Heisenberg model with positive exchange interaction,
\begin{equation}\label{}
H = J \sum_{(i,j)}\vec{ S}_{i}\cdot \vec{S}_{j}, \qquad \qquad \qquad (J > 0),
\end{equation}
the effects of quantum fluctuations were poorly understood.
Most notably, the one-dimensional $S=1/2$ model had been solved
exactly by 
Bethe in 1931\cite{bethe}, and in his {\sl Bethe Ansatz} solution, it
was clear there was
no long range order, indicating that at least in one dimension, 
quantum fluctuations overcome the long-range order. This issue worried Landau so
much, that by the 1940's he had abandoned the idea of 
antiferromagnetism in quantum spin systems\cite{pomeranchuk}.
Phil Anderson reflects on this uncertainty in his 1952 article ``An Approximate Quantum Theory of the
Antiferromagnetic Ground State''
\cite{1952},
writing \begin{quotation}
{\sl ``For this reason the very basis for the recent
theoretical work which has treated antiferromagnetism  similarly to
ferromagnetism remains in question. In particular, since the
Bethe-Hulth\'en ground-state is not ordered, it has not been certain
whether an ordered state was possible on the basis of simple
$\vec{S}_{i}
\cdot \vec{S}_{j}
$ interactions''}
\end{quotation}

The  situation 
began to change in 1949,
with Shull and Smart's\cite{shullsmart} detection of antiferromagnetic order in 
MnO by neutron
diffraction, which 
encouraged Anderson to turn to the unsolved
problem of zero-point motion in
antiferromagnets. 
Early work on spin-wave theory
by Heller, Kramers and Hulth\'en 
had treated spin waves as classical excitations, but
later 
work by 
Klein and Smith\cite{kleinsmith} had noted that 
quantum zero point motions
in a spin $S$ ferromagnet correct the ground-state energy by an 
amount of order $1/S$, a quantity that becomes increasingly small as
the size of the spin increases. 
It is this effect that increases the ground-state
energy of a ferromagnet from  its classical value 
$E\propto - J \langle {\vec{S}^{2}}\rangle = -JS(S+1)$ to its exact quantum value
$E\propto -J S^{2} $. 

\fight=\textwidth
\fg{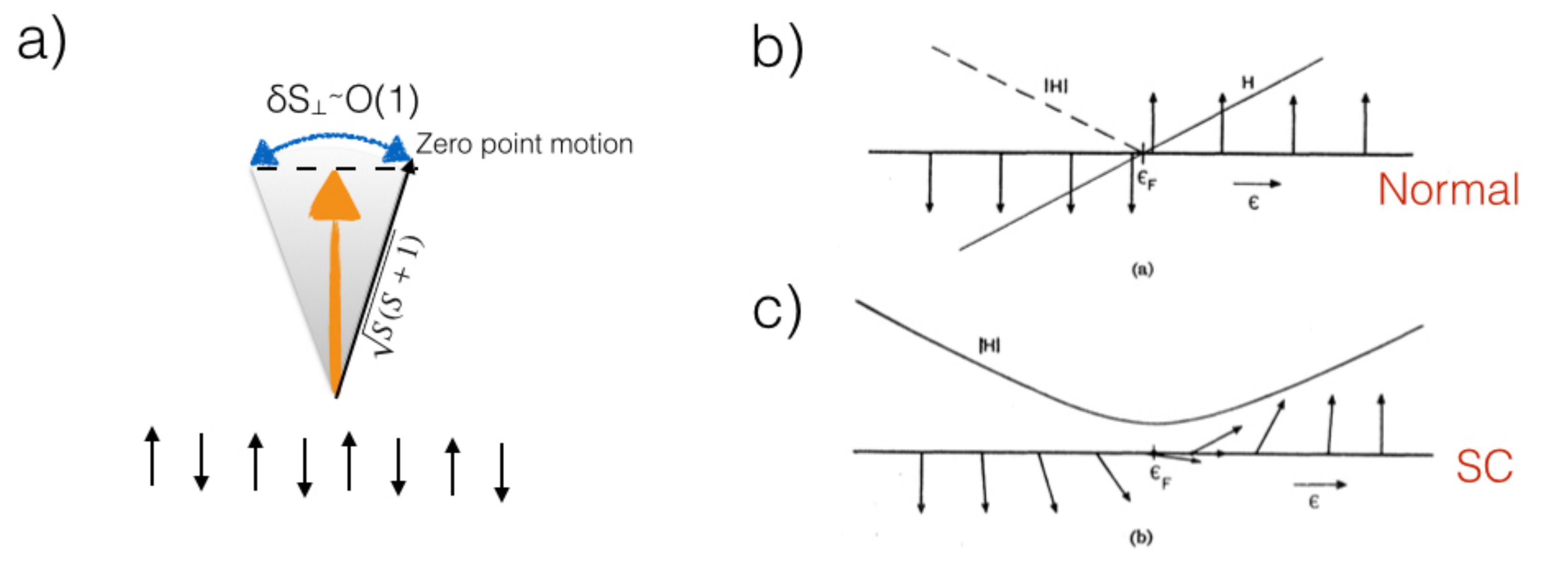}{fig2}{(a) Reduction of staggered moment from length
$\sqrt{S (S+1)}$ to semi-classical value $\sqrt{S (S+1)}- O (1)$. (b) In
Anderson's domain wall interpretation of superconductivity
the normal
Fermi liquid is a sharp domain wall
where the Weiss field $H$ vanishes at
the Fermi surface; (c)
in the superconductor the
pseudospins rotate smoothly and Weiss field never vanishes, giving
rise to a finite gap. 
}

A key result of Anderson's work is an explanation for the survival 
of antiferromagnetic 
order in two and higher dimensions, despite its absence
in the Bethe chain. His 
expression for the reduced
sublattice magnetization (Fig. 2a) of a bipartite antiferromagnet,
is
\begin{eqnarray}\label{l}
\langle S_{z}\rangle &=&\sqrt{S(S+1)- \delta S^{2}_{\perp }}\cr &=& S \left[1-
\frac{1}{2S}\int \frac{d^{d}q}{(2 \pi
)^{d}}
\left(
\frac{1}{\sqrt{1-\gamma_{\bq }^{2}}} -1\right) + 
O (\frac{1}{S^{2}}) \right]
\end{eqnarray}
where $\gamma_{\bq }= \frac{1}{d}\sum_{ l =1,d}\cos q_{l}$.  A similar
result was independently discovered by Ryogo Kubo\cite{kubo}. 
In an antiferromagnet, 
the staggered magnetization does not
commute with the Hamiltonian and thus undergoes continuous zero point
fluctuations that reduce its magnitude (Fig. 2a). 
Since $\sqrt{1-\gamma_{\bk }^{2}} \sim |\bq|$ at small
wavevector $\bq $, these fluctuations become particularly intense at
long wavelengths, with a reduction in magnetization 
\begin{equation}\label{}
\Delta M  \sim \int \frac{d^{d}q}{q}\sim \left\{\begin{array}{cc}
\infty & (d=1)\cr
\hbox{finite} & (d\geq 2)
\end{array} \right.
\end{equation}
In this way, Anderson's model calculation 
could account for the absence of long range order in the Bethe
chain as a result of long-wavelength quantum fluctuations and the
stability of antiferromagnetism in higher dimensions. 

At several points in his paper, Phil  muses on
the paradox that the ground-state of an antiferromagnet is 
a singlet, with no preferred direction, 
a thought he would return to 
in his later work on resonating valence bonds with Patrick
Fazekas\cite{fazekas}.  For the moment however, Phil resolves
the paradox by estimating
 that the time for an antiferromagnet to invert its spins by 
tunnelling is macroscopically long, 
so that the sublattice magnetization becomes an observable
classical quantity. 
Phil's semi-classical treatment of the
antiferromagnet would later set the stage
for Duncan Haldane\cite{haldane82} to carry out a semi-classical 
treatment of the one dimensional
Heisenberg model, revealing an unexpected topological term.
But in the near future, Anderson's study of antiferromagnetism 
had influence in a wholly unexpected direction: superconductivity. 

One of the issues that was poorly understood 
following the Bardeen Cooper Schrieffer (BCS) theory of
superconductivity, 
was the question  of charge fluctuations. 
In an insulator the charge gap leads to a 
dielectric with no screening. However, were the superconducting gap 
to have the same effect, it would eliminate the weak
electron phonon attraction. It was thus
essential to show that both screening and the longitudinal plasma mode
survive the formation of the BCS gap. 
In his 1958 paper ``Random-Phase Approximation in the Theory of
Superconductivity'' 
\cite{pwarpa}, Phil  writes
\begin{quotation}
{\sl ``Both for this reason, and because it seems optimistic to assume
that the collective [charge fluctuation] and screening effects (which
are vital even in determining the phonon spectrum) will be necessarily
unaffected by the radical changes in the Fermi sea $\dots $, it is
desirable to have a theory of the ground-state of a superconductor
which can simultaneously handle these collective effects in the best
available approximation 
$\dots $''}
\end{quotation}

Phil's experience with antiferromagnetism enabled him to make a
new link between magnetism and superconductivity. He observed that if one considered
a pair to be a kind of ``down-spin'' and the absence of a pair to be a
kind of ``up spin'' in particle hole space, 
\begin{eqnarray}\label{l}
\hbox{no pair:}\qquad \qquad 
\vert \Uparrow\rangle &\equiv&\vert 0\rangle, \cr
\hbox{pair:}\qquad \qquad \vert \Downarrow
\rangle &\equiv & c\dg_{k\uparrow}c\dg_{-k\downarrow }
\vert 0 \rangle
\end{eqnarray}
then the BCS ground-state is revealed as a kind of 
Bloch domain wall (Fig. 2b,c) 
formed around the Fermi surface\cite{pwarpa}.   
This new interpretation  forges a link between 
between superconductivity and antiferromagnetism, 
enabling the pairing field to be identified
as a transverse Weiss field in particle-hole space.  
Moreover the analogy works at a deeper level, because 
like in quantum antiferromagnetism, the superconducting order parameter is
non-conserved, allowing it to fluctuate and importantly, to 
deform in response to an electric field, preserving the screening.

Let us look at this in a little more detail. BCS theory involves three key operators, the number operator $
(n_{k\uparrow}+ n_{-k\downarrow })$, the pair creation and pair
annihilation operators, $b\dg_{k} = c\dg_{k\up}c\dg_{-k\dw}$, 
$b_{k}= c_{-k\dw}c_{k\up}$.  The key observation was to identify these
operators as the components of a pseudo-spin.
In the subspace where $n_{k\uparrow}+n_{-k\dw}$ is either $0$or $2$,
Anderson's defined the pseudospin as 
\begin{equation}\label{}
2s_{z}= 1 - n_{k}- n_{-k} = 
\pmat{1&0\\ 0&-1},
\end{equation}
so that a fully  occupied $k$ state is a ``down''  pseudo-spin, and an
empty $k$-state is an ``up'' spin.  Similarly, the 
the raising and lowering operators are respectively, 
the pair destruction and
creation operators 
\begin{equation}\label{}
b_{k} \equiv s_{xk}+is_{yk}
=\pmat{0&0\\ 1 & 0}, 
qquad 
b\dg_{k}\equiv s_{xk}-is_{yk} =  \begin{pmatrix} 0 & 0 \\ 1 & 0 \end{pmatrix}.
\end{equation}
In this language, the BCS reduced Hamiltonian 
\begin{equation}\label{}
{\cal H}_{\hbox{RED}} = - 2 \sum_{k}\epsilon_{k}s_{zk}- \sum_{k,k'}
V_{k,k'} \vec{ s}_{\perp k}\cdot \vec{s}_{\perp k'}
\end{equation}
is a kind of magnet that resides in momentum space.  Anderson showed
that in this language, the metal is a sharp
domain wall along the Fermi surface (Fig. 2b) while the superconductor has a
soft ``Bloch domain wall'' (Fig 2c) in which the pseudo-spins rotate
continuously from  down (full) to sideways (linear combination of full
and empty) to up (empty). By calculating the spin-wave fluctuations
Anderson was able to show that with the Coulomb interaction included,
the longitudinal electromagnetism of the metal, its screening and
plasma modes, are unaffected by the superconducting gap.

Anderson's pseudo-spin reformulation of BCS had a wide influence. 
Two years later,
Nambu extended the pseudo-spin 
approach to reformulate Gor'kov's Green function
approach using his now famous ``Nambu matrices''\cite{nambu}.  Perhaps most
important of all, by making the analogy between superconductivity and
magnetism, the community took a cautious step closer to regarding the
superconducting phase as a palpable, detectable variable (with the
caveats of gauge fixing). This
new perspective, especially the link
between phase, supercurrents and gauge invariance, would 
soon culminate in
Anderson's ideas on how gauge particles acquire mass - the {\sl
Anderson Higgs} mechanism (see Witten's article in this volume). 



\section{Orange Period: Moment Formation and the Kondo Problem}\label{}

\subsection{Superexchange}\label{}

Towards the end of the 1950's, Anderson began to turn his attention
towards the microscopic origins of antiferromagnetism. In his 
1959 paper ``New Approach to the Theory of
Superexchange Interactions''\cite{PhysRev.115.2} 
Phil argues that origin of antiferromagnetism is
{\sl the Mott mechanism}, i.e the Coulomb cost of doubly occupied orbitals.
Phil 
writes 
\begin{quotation}
{\sl ``In such a simple model all the degenerate states in the
ground-state manifold have exactly one electron per ion, while all
the excited states with one transferred electron have energy
$U$. Between an pair of ions at a distance ${\bf R}-{\bf R}'$ there is
only one (hopping term) $b_{\bR-\bR'}$; this must act to return
the state to one of the ground manifold... so that 
\begin{equation}\label{}
\Delta E = \hbox{constant}+ \sum_{\bR,\bR'} \frac{2 \vert
b_{\bR-\bR'}\vert^2}
{U}\bS_{\bR}\cdot \bS'_{\bR'},
\end{equation}
This is the antiferromagnetic exchange effect.''}
\end{quotation}
This paper contains the origins of our modern understanding
of Mott insulators, including an early formulation of the Hubbard
model with Anderson's hallmark use of {\sl U} to denote the onsite 
Coulomb repulsion.

\subsection{Anderson's model for local moment formation}\label{}

While the notion of local moments is rooted in 
early quantum mechanics, the mechanism of moment formation
was still unknown in the 1950s. 
At this time, experiments at Orsay
and at Bell Labs started to provide valuable new
insights. In Orsay, 
Jacques Friedel
and Andr\'e Blandin proposed that virtual bound-states
develop around localized d-states in a metal, arguing that 
ferro-magnetic exchange forces then split these resonances to form local
moments.  
Recalling the first time he encountered this idea, 
Phil writes\cite{kondoII,career}:
\begin{quotation}
{\sl In the Fall of '59, a delightful little discussion meeting on
magnetism in metals was held in Brasenose College, Oxford. ...
Blandin presented the idea of virtual states and I introduced 
the conceptual
basis for antiferromagnetic s-d exchange, without any understanding,
at least on my part, that the two ideas belonged together. The only
immediate positive scientific result of the meeting was that I won a
wager on the sign of the Fe hyperfine field on the basis of these ideas.''}
\end{quotation}

Around this time, 
Bernd Matthias's group at  Bell Labs
discovered that
the development of a localized moment on iron atoms 
depends on the metallic 
environment - for example, iron impurities dissolved in 
niobium do not develop a local moment, yet they do so in the
niobium-molybdenum alloy, Nb$_{1-x}$Mo$_{x}$ once the concentration of
molybdenum exceeds 40\% ($x>0.4$). 
  Anderson was
intrigued by this result and realized that while it was probably connected to the virtual
bound-state ideas of Friedel and Blandin, 
ferromagnetic exchange was 
two weak to drive moment formation.  Once again, he turned to 
the  Mott mechanism as a driver and 
the key element of his theory ``Localized Magnetic States in
Metals'',  is the repulsion
between {\sl anti-parallel} electrons in the same orbital, given by the
Coulomb repulsion integral,
\begin{equation}\label{ueq}
U = \int \vert \phi_{loc} (1)\vert^{2}e^{2}r_{12}^{-1}\vert \phi_{loc} (2)\vert^{2}d\tau.
\end{equation}
Phil emphasizes this point, writing 
\begin{quotation}
{\sl ``the formal theory is much more straightforward if one includes U in the manner in which we do it, as a repulsion of opposite-spin electrons in $\phi_{loc}$, not as an attraction of parallel ones''}
\end{quotation}
Another new element of Anderson's theory of moment formation, not
contained in earlier theories, was the explicit formulation 
of his model as a quantum field theory. 
The heart of the Anderson model is a hybridization term 
\begin{equation}\label{}
H_{sd} = \sum_{\bk ,\sigma }V_{d\bk } (c_{\bk \sigma }\dg c_{d\sigma
}+ c\dg_{d\sigma }c_{\bk \sigma }),
\end{equation}
which mixes s and d electrons, generating a virtual bound-state of
width
$\Delta = \pi \langle V^{2}\rangle \rho $, where $\rho (\epsilon)$ is
the conduction electron density of states and $\langle V^{2}\rangle $
the Fermi surface average of the hybridization, and 
the onsite Coulomb interaction, 
\begin{equation}\label{}
H_{corr}= U n_{d\uparrow}n_{d\downarrow },
\end{equation}
where $U$ is as given in (\ref{ueq}).   
With these two terms, Phil
unified the Freidel-Blandin virtual bound-state resonance with the 
``Mott mechanism''  he had already introduced for insulating
antiferromagnets.   Using a mean-field 
Hartree-Fock treatment of his model, Anderson shows that 
if 
\begin{equation}\label{}
U> \pi\Delta 
\end{equation}
the virtual bound state resonance splits into two. 
One of the aspects of the paper that may have been confusing at the
time, was that taken literally, this suggested a real phase transition
into a local moment state. 
Anderson clearly did not see it this way, 
\begin{quotation}
{\sl ``It is the great conceptual simplification of the impurity
problem that is possible to separate the question of the existence
of the ``magnetic state'' entirely from the actually irrelevant
question of whether the final state is ferromagnetic,
antiferromagnetic or paramagnetic.''}
\end{quotation}
Today we understand that 
Anderson's mean-field description of magnetic moments captures
the physics at intermediate time scales, describing a
cross-over in the renormalization trajectory as it makes a close fly-by
past the repulsive local moment captured in Phil's mean
field theory. 

\subsection{The Kondo Model}\label{}

A central prediction of Phil's 1961 paper was that
the residual interaction between the local moment, and the surrounding
electrons, the {\sl s-d} interaction
is {\sl antiferromagnetic}.   
By freezing the local
moment,  Phil was able to calculate the small shifts in the conduction
electron energies, demonstrating they were indeed antiferromagnetic.
He writes 
\begin{quotation}
{\sl `` Thus any g shifts caused by free electron polarization will
tend to have antiferromagnetic sign.''}
\end{quotation}
Phil had of course guessed this in his 1959 bet at Brasenose College
Oxford!

The conventional wisdom of the time expected a ferromagnetic s-d
exchange. Indeed,  
the {\sl s-d model} describing the interaction of local
moments with conduction electrons had been formulated 
by Clarence Zener\cite{zenersd}
and written down in second-quantized notation in the 1950s, 
by Tadao Kasuya\cite{kasuya1956}, but with a {\sl ferromagnetic} 
interaction, derived from exchange.
\footnote{Curiously, in his 1961
paper, Phil does not mention super-exchange as the origin of the
antiferromagnetic s-d interaction, despite his development of this
idea in his 1959 paper on insulating antiferromagnets. 
Perhaps it was felt that
metals are different. It was not until
the work of Schrieffer and Wolff\cite{swolf} that the Kondo interaction was
definitively identified, using a careful canonical transformation, 
as a form of super-exchange interaction, of magnitude
\begin{equation}\label{}
J \sim \frac{4\langle V_{d\bk }^{2}\rangle }{U}.
\end{equation}
}

On the other side of the world in Tokyo,
one person, Jun Kondo realized
that Anderson's  prediction of an antiferromagnetic s-d coupling
would have experimental consequences, and his efforts to reveal them 
led to the solution of a 30
year old mystery. Following Anderson's prediction, 
Kondo now wrote down a simple model for the 
{\sl antiferromagnetic }
interaction, 
\begin{equation}\label{}
H_{K} = \sum \epsilon_{k}c\dg_{\bk \sigma }c_{\bk \sigma } +  
J \vec{S}_{d}\cdot \vec{\sigma } (0), 
\end{equation}
where $\vec{\sigma } (0)$ is the electron spin density at the site of the
magnetic moment, 
and he set out to examine the consequences of the  antiferromagnetic
exchange. 
This led Kondo to 
calculate the magnetic scattering rate 
$\frac{1}{\tau }$ of electrons to cubic order in the s-d interaction. 
To his surprise, the cubic term contained 
a logarithmic temperature dependence\cite{kondo2}:
\begin{equation}\label{kondotau}
\frac{1}{\tau } \propto \left[J\rho + 2 (J\rho )^{2}\ln \frac{D}{T} \right]^{2}
\end{equation}
where $\rho $ is the density of states of electrons in the conduction
sea and $D$ is the half-width of the electron band. 
Kondo noted that if the s-d
interaction were positive and antiferromagnetic, 
then as the temperature is lowered, the coupling constant, 
the scattering rate and resistivity start to rise. This meant that 
once the magnetic scattering overcame the phonon scattering, the resistance
would develop a resistance minimum. 
Such resistance minima had been seen in metals for more than 30
years\cite{resistanceminimum,resistanceminimum2}. Through Kondo 
and Anderson's work, this thirty year old mystery 
could be directly interpreted as a 
a direct consequence of the predicted antiferromagnetic s-d
interaction with local magnetic moments.

\subsection{Kondo's result poses a problem}\label{}

After Kondo and
Anderson's work, the community quickly realized that the 
``Kondo effect'' raised a major difficulty. 
You can see from (\ref{kondotau})
that at the ``Kondo temperature'' $T\sim T_{K}$ 
where $ 2J\rho \ln (D/T_{K})\sim 1$,  or
\begin{equation}\label{}
T_{K}\sim D e^{-1/(2J\rho)}
\end{equation}
the Kondo log becomes comparable
with the bare interaction, so that at lower temperatures
perturbation theory fails. 
What happens at lower temperatures once perturbation theory fails?  
This is the ``Kondo problem''.

By the late 1960's, from  the work of early pioneers on the Kondo problem,
including Alexei Abrikosov, Yosuke Nagaoka, Harry Suhl, Bob Schrieffer
and Kei Yosida,  much had been learned about the Kondo problem.  It had
become reasonably clear that at low temperatures the Kondo coupling
constant grew to strong coupling, to form a spin singlet, but the
community was divided over whether the residual scattering would be
singular, or whether it would be analytic, forming an ``Abrikosov
Suhl'' resonance. The problem also lacked a conceptual
framework and there were no controlled approximations.

\subsection{How a Catastrophe led to new insight}\label{}

The solution to
Kondo problem required a new concept - the
renormalization group. 
Today we know the Kondo effect as an example of asymptotic freedom
- a running coupling constant that flows from weak coupling at high
energies, to strong coupling at low energies, ultimately binding the
local moment into a singlet with electrons in the conduction sea. 
In the late 60's, renormalization had
entered condensed matter physics as a new tool for statistical
physics. Phil and his collaborators now brought 
the renormalization group to quantum condensed matter by 
mapping the Kondo problem onto a one-dimensional Ising model with long
range interactions. 

Phil entered the field from an unexpected direction after discovering 
an effect known as the
{\sl orthogonality} or {\sl X-ray} catastrophe. 
Phil's 1967 paper ``Infrared Catastrophe in Fermi Gases with
Local Scattering Potentials''\cite{xraycat}, was stimulated
by a conversation with John Hopfield, who speculated that the 
introduction of an impurity potential into a Fermi gas 
produces a new ground-state  $\vert \phi^{*}\rangle $
that is orthogonal to the original ground-state $\vert \phi_{0}\rangle $. 
Phil examined this idea in detail, and 
showed  that
when a local scattering potential suddenly changes, in the
thermodynamic limit, the overlap between the original and the new Fermi gas ground-states 
identically vanishes $\langle \phi_{0}\vert  \phi^{*}\rangle =0$. 
For example, when an X-ray ionizes an atom in a
metal, the ionic potential suddenly changes and this causes the
conduction sea to evolve 
from its original ground-state $\vert \phi_{0}\rangle$
into a final-state $\vert \phi_{f} (t)\rangle  = e^{-i Ht}\vert
\phi_{0}\rangle $\cite{Mahan:1967dc}. 
In fact, Phil showed that the resulting 
relaxation is {\sl critical}, with power-law decay in the overlap amplitude
\begin{equation}\label{}
G (t)= \langle \phi_{0}\vert \phi_{f} (t)\rangle
\sim
\frac{e^{-i \Delta E_{g}t}}{t^{\epsilon}},
\end{equation}
where $\Delta E_{g}$ is difference between final and initial
ground-state energies. The absence of a
characteristic time-scale indicates that the relaxation into the
final-state ground-state is infinitely slow. 
By Fourier transform, this implies a singular density of
states\cite{Mahan:1967dc} \begin{equation}\label{}
\rho (E)\sim \int dt G (t)e^{iEt} \sim (E-E_{g})^{-1+\epsilon}.
\end{equation}
This topic was also studied by Mahan\cite{Mahan:1967dc} who linked the subject with X-ray
line-shapes. 
Nozi\` eres and de Dominicis\cite{xray2} later found an exact solution to the
integral equations of the orthogonality catastrophe. 
The X-ray catastrophe is also responsible for the 
singular Green's functions of electrons in a one-dimensional Luttinger
Liquid\cite{andersonirc,Imambekov:2009gi,Fiete:2009gp}. 

One of the key conclusions of this work 
was that the orthogonality catastrophe {\sl occurs} in the Kondo
problem.  Phil recognized
that each time a local moment flips, 
the Weiss field it exerts on conduction electrons reverses, 
driving an orthogonality catastrophe in the ``up'' and
``down'' electron fluids. 
In its 
anisotropic form, the Kondo interaction takes the form
\begin{equation}\label{}
H_{K}= J_{z} S_{dz} \sigma_{z} (0)+ J_{\perp }\left[S_{+}\sigma_{-}
(0) + S_{-}\sigma_{+} (0) \right]
\end{equation}
where $\sigma_{\pm }= (\sigma_{x}\pm i \sigma_{y})/2$ and $S_{\pm } =
S_{dx}\pm i S_{dy}$ are the local lowering and raising operators for the
mobile conduction and localized d-electrons respectively.  
From the work of Nozi\`eres
and de Dominicis, the amplitude for two  spin flips at times $t_{1}$ and $t_{2}$
is  
\begin{eqnarray}\label{l}
(J_{\perp })^{2}\left(\frac{\tau_{0}}{t_{2}-t_{1}}
\right)^{2-2\epsilon} = (J_{\perp })^{2}\exp \left[-(2-\epsilon)\ln  \left(\frac{t_{2}-t_{1}}{\tau_{0}} \right)\right].
\end{eqnarray}
where $\tau_{0}$ is the short-time (ultra-violet) cut-off and 
$\epsilon \sim 2J_{z}\rho $ is
determined by the change in the scattering phase shift of the up and
down Fermi gases, each time the
local spin reversed. This suggested that the quantum spin flips in a 
Kondo problem could be mapped onto the statistical mechanics of 
a 1D Coulomb gas of ``kinks'' with a logarithmic interaction. 

\subsection{The Anderson-Yuval solution to the Kondo problem}\label{}

Working with graduate student Gideon Yuval\cite{yuval,yuval2} and a little later, Bell Labs
colleague  Don Hamann,  
Phil's team took up the task of organizing
and summing 
the X-ray divergences of multiple spin-flip processes as a continuous time
path-integral. 
With some considerable creativity, it became
possible to map the quantum partition function of the Kondo model onto the
{\sl classical } partition function of a Coulomb gas
of kinks. By regarding the kinks as domain walls in a one dimensional
Ising model, they could further map the problem onto 
a  one-dimensional Ising Ferromagnet
spin chain with a  $1/r^{2}$ interaction, 
\begin{equation}\label{}
\frac{H}{T} = - (2-\epsilon)\sum_{i>j}\frac{S^{z}_{i}S^{z}_{j}}{|i-j|^{2}} - \mu\sum_{i} S^{z}_{i}S^{z}_{i+1},
\end{equation}
where the Ising spins can have values $S_{j}=\pm 1/2$ at each site;
the position $j$ along the chain 
is really the imaginary time $\tau  = j \tau_0$ measured in units of the 
short-time cut-off, with periodic boundary conditions and a total
length determined by the inverse temperature, $L = \frac{\hbar
\tau_{0}}{k_{B}T}$.  
The tuning 
parameter $\epsilon=J_{z}\rho $ is determined by the Ising part of the
exchange interaction,  while the transverse interaction $J_{\perp }$
sets
$\mu = - 2 \ln  J_{\perp }\rho $, 
the chemical potential of 
domain-wall kinks in the ferromagnetic spin chain. The larger
$J_{\perp }\rho $, the more kinks are favored.

Suddenly a complex quantum
problem became a tractable statistical mechanics model. 
It meant one could adapt the
renormalization group from statistical physics to examine how the
effective parameters of the Kondo parameter changed at lower and lower
temperatures. By integrating out the effects of two closely
separated pairs of spin flips, Anderson, Yuval and Hamann\cite{Anderson:1970bv} derived the
scaling equations
\begin{eqnarray}\label{l}
\frac{\partial J_{z}}{\partial \ln \tau_{0}}&=& J_{\perp }^{2}, \cr
\frac{\partial J_{\perp }}{\partial \ln \tau_{0}}&=& J_{\perp }J_{z}.
\end{eqnarray}
Under
these scaling laws, $J_{\perp }^{2}-J_{z}^{2}$ is conserved, giving
rise to the famous scaling trajectories shown in Fig. \ref{XXXX}. 
There are two fixed points:
\begin{enumerate}

\item $\epsilon\sim J_{z}\rho <0 $
Ferromagnetic ground state $\equiv $ unscreened local moment.

\item $\epsilon \sim J_{z}\rho >0$
``Kink liquid'' $\equiv $ screened local moment, where the Kondo
temperature sets the typical kink separation $l_{0}\sim T_{K}$.

\end{enumerate}
\fg{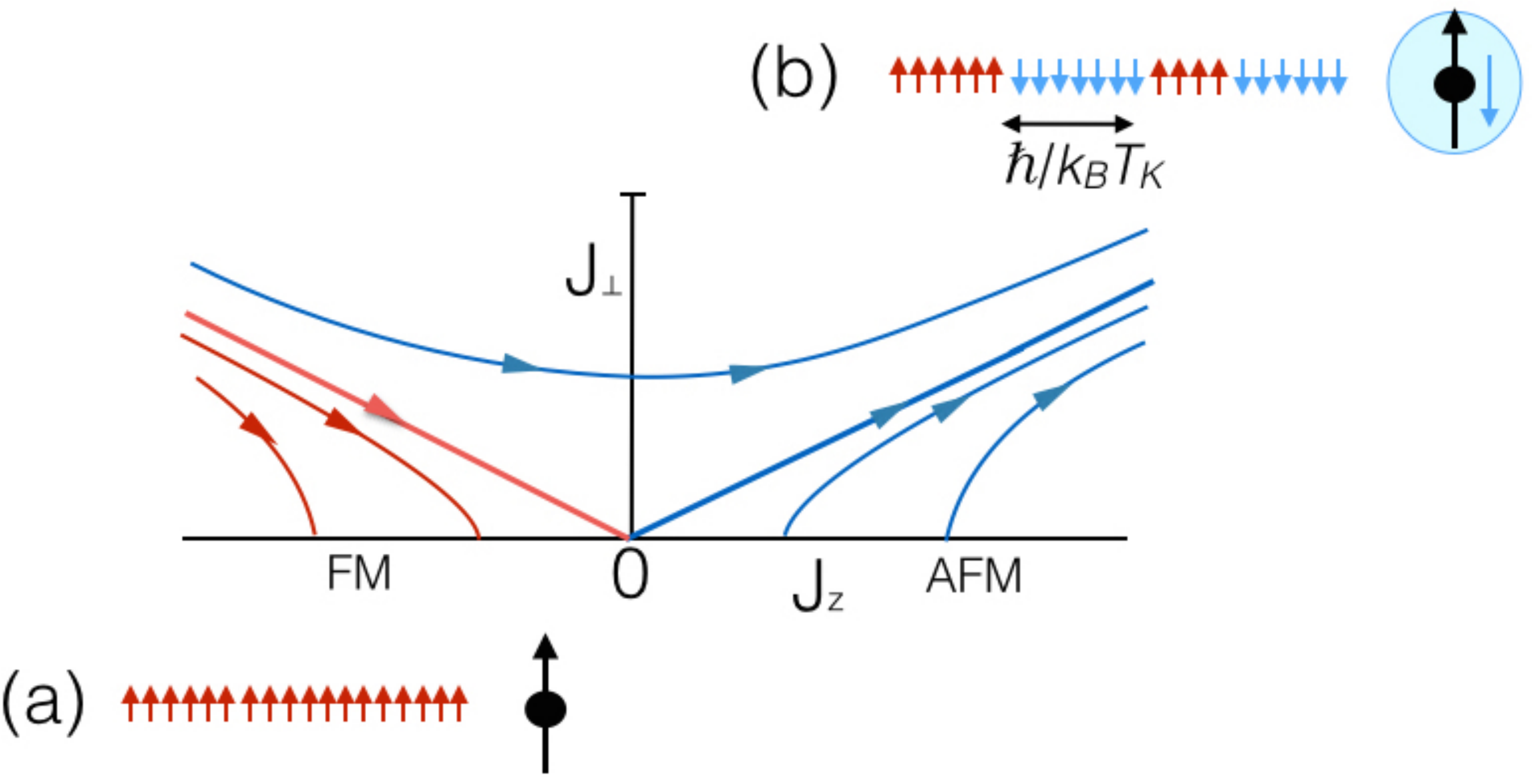}{XXXX}{The Anderson-Yuval-Hamann scaling curves\cite{Anderson:1970bv} for
the Kondo model. (a) Red trajectories correspond to an unscreened moment,
or a kink-free Ferromagnetic ground-state in the magnetic
analogy. (b) Blue trajectories correspond to a fully screened local
moment, or a ``kink condensate'' where the average kink separation is
determined by the inverse Kondo temperature.}

In this way, the Coulomb gas of 
kinks had a phase transition at $J_{z}=0$. 
For negative $J_{z}<0$, the kinks are absent, but for positive 
$J_{z}$, the chemical potential of the kinks grows so
that they proliferate, forming a kink-liquid.  Although
Anderson, Yuval and Hamann were unable to completely solve the strong
coupling problem, the problem was solvable for the so called {\sl Toulouse
limit}, where $\epsilon=2J_{z}\rho =1$, and in this limit, it could
be shown that the strong coupling limit was free of any singular
scattering. In their paper\cite{Anderson:1970bv}, the authors conclude that 
\begin{quotation}
{\sl ``
The most interesting question on the Kondo effect has been from the start
whether it did or did not fit into the structure of usual Fermi gas
theory: In particular, does a true infrared singularity occur as in
the x-ray problem, or does the Kondo impurity obey phase-space
arguments as $T\rightarrow 0$ and give no energy dependences more
singular than $E^{2}$(or $T^{2}$), and is  [the susceptibility] $\chi
(T=0)$ finite?  The result we find is that the usual antiferromagnetic
case in fact \underline{does} fit after the time scale has been revised
to $\tau_{\kappa}$, i.e. that it behaves like a true bound-singlet as
was conjectured originally by Nagaoka. 
"}
\end{quotation}
i.e the authors conclude that ground-state of  the Kondo problem is a Fermi liquid.

During this period, Phil wrote
series of informal papers in Comments in Solid State
Physics\cite{kondoI,kondoII,kondoIII,kondoIV} that provided a very
personalized update on the progress. The
last of these papers, ``Kondo effect IV: out of the wilderness''\cite{kondoIV}, 
summarizes what become the {\sl status quo} in this
problem. Phil writes
\begin{quotation}
{\sl ``In conclusion then, the status is this: we understand very clearly
the physical nature of the Kondo problem, which is beautifully
expressed in Fowler's picture of scaling: electrons of high enough
energy interact with the weak, bare interaction and the bare Kondo
spin, but as we lower the energy the effects of the other electrons
gradually strengthen the effective interaction until finally, at
energies near $T_{K}$, the effective interaction starts to get so
large that we must allow the local spin to bind a compensating spin to
itself, and the Kondo spin effectively disappears, being replaced by a
large resonant nonmagnetic scattering effect. My own opinion is that
the low temperature behavior is totally non-singular, the Kondo
impurity looking simply like a localized spin fluctuation site, but
others believe that there may remain a trace of singular behavior.''
}
\end{quotation}
The influence of these ideas ran far and wide:
\begin{enumerate}

\item It introduced scaling theory to quantum systems.  The project
started by Anderson and Yuval later culminated in 
Wilson's numerical renormalization work\cite{revwilson}.  
\fg{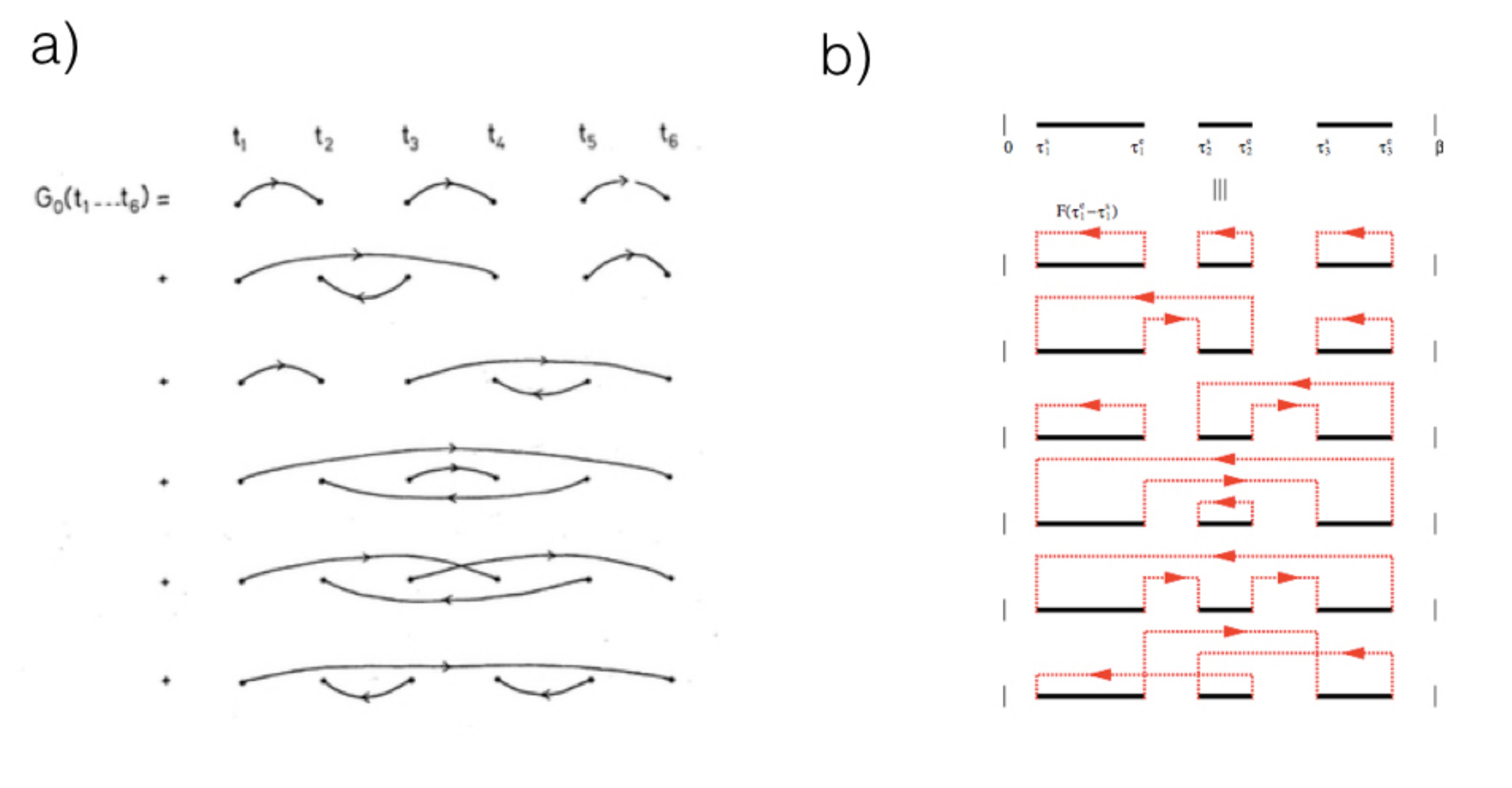}{fig4}{
Figure contrasting the use of the
Yuval-Anderson approach to impurity models\cite{yuval} with the modern continuous
time Monte Carlo methods\cite{millisref}: (a) illustrative sixth order diagram
from the original Anderson-Yuval paper\cite{yuval}, (b) the same 
set of diagrams used in
\cite{millisref}. The black arrowed lines in (a) and the red-arrowed 
lines in (b) describe conduction electrons propagating between spin
flips. 
}

\item The key conclusion about the non-singular character of the
ground-state, confirmed by Wilson, later became the basis for
Nozi\` eres'
strong coupling treatment of the key low temperature
properties of the Kondo problem\cite{noziereskfl}, which by mapping
the physics onto a local Fermi liquid,  accounted for the
two-fold enhancement of the Sommerfeld-Wilson ratio observed in
Wilson's numerical work\cite{revwilson}.

\item The statistical mechanics of the problem, with scale-dependent 
interactions between topological defects, provided inspiration and 
ground-work for Kosterlitz and Thouless's scaling
solution\cite{Kosterlitz:1973fc} to the Berezinski-Kosterlitz-Thouless transition in the 2D xy
antiferromagnet, whose scaling flows replicate the Anderson-Yuval-Hamann diagram.

\item Phil's belief that Kondo model would be exactly solvable was
dramatically confirmed by the independent Bethe Ansatz solutions of
Natan Andrei and Paul Weigman\cite{PhysRevLett.45.379,Vigman:1980uc} 

\item Modern continuous-time Montecarlo solvers for 
computational
Dynamical Mean Field Theory approaches to materials research are a direct
descendant of the Anderson-Yuval mapping of quantum impurity models  to
statistical mechanics in time (see Fig. \ref{fig4}
)\cite{millisref,hauleetalref}. 

\end{enumerate}

\section{Green period: Mixed valence and the large $N$ expansion. }

The solution to the Kondo problem resurrected 
the question about when fluctuations (quantum, random)
defeat anti-ferromagnetic order,  
and when they do, what replaces it?  Leaving 
the Kondo wilderness behind, Anderson's magnetic life developed in directions
that explore this question.
One way to avoid magnetism is by enhancing zero-point
fluctuations with frustration, 
as in a triangular lattice antiferromagnet, and it was this direction 
Phil explored with Patrik Fazekas\cite{fazekas}, leading them 
to apply Pauling's
resonating valence bond idea to spin liquids (see Fig. \ref{newdirs} a). 
A second way is through
quenched disorder, as in dilute magnetic alloys, and this led Phil,
with Sam Edwards to invent the concept of the spin
glass\cite{Anderson1970sg,edwardsanderson} (see Fig. \ref{newdirs} b).  A third direction, 
is through quantum fluctuations
induced by the Kondo effect and valence fluctuations.  This returned
Anderson to the unfinished business of the Anderson and Kondo lattices
(see Fig. \ref{newdirs} c). 
The first
two directions are discussed in the excellent articles by Ted
Kirkpatrick, Patrick Lee and Mohit Randeria in this volume. Here,  I will
focus the discussion on Phil's contributions to our understanding of
the Kondo lattice and valence fluctuations.

In the 1970's experimentalists started to investigate 
the fate of dilute magnetic alloys 
as the magnetic atoms become
more concentrated. In transition metal alloys, 
the RKKY interaction between the magnetic ions
overcomes the Kondo effect, giving rise to spin glasses
\cite{Anderson1970sg,Cannella:1972ea}. But in
rare earth and actinide intermetallic compounds, 
the Kondo effect and associated valence fluctuations are strong enough
to overcome the magnetism, even in fully concentrated lattices of
local moments,  leading to a wide variety of {\sl heavy
fermion} materials. 
Phil's insights played a vital role in the
development of the field.
\fg{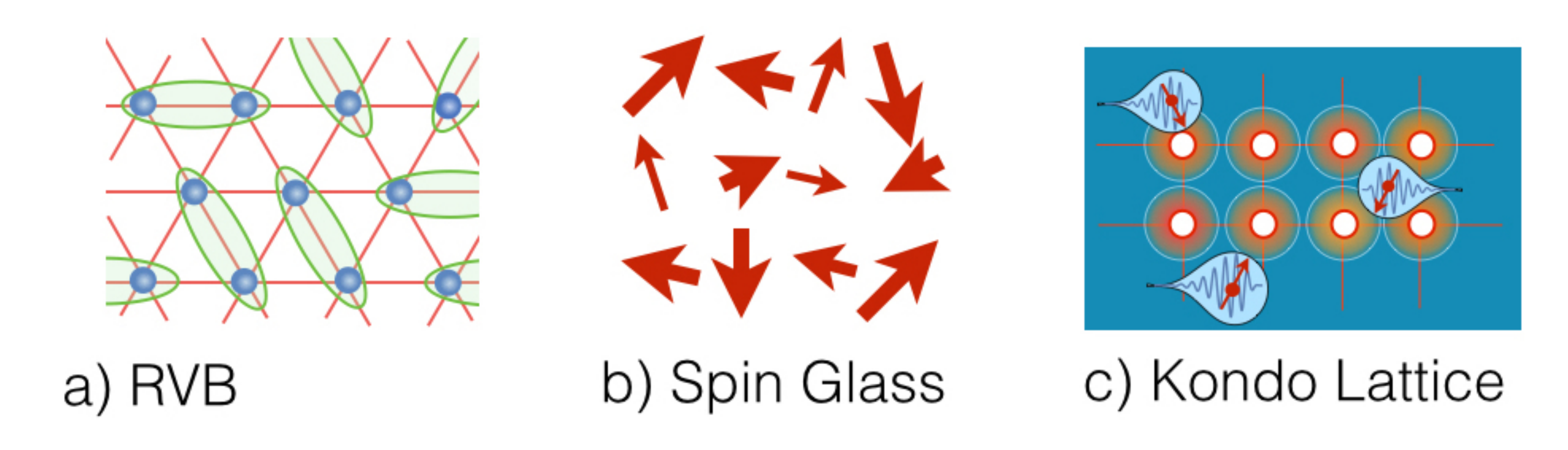}{newdirs}{Three new directions for Anderson's research in
the 1970s: a) the resonating valence bond ground state for the
frustrated triangular lattice \cite{fazekas}, b) the spin-glass
ground-state for a frustrated disordered array of
spins\cite{edwardsanderson} and c) the problem of mixed valence, where
mobile heavy electrons move through a lattice of Kondo screened local
moments.}

Early experimental progress in the new field of {\sl mixed valence} was rapid
and chaotic.  A plethora of new 
intermetallic compounds
were discovered which display local moment physics at high
temperatures, but which instead of magnetically ordering at low
temperatures, form an alternative ground-state. 
Already in 1969, the group of Ted
Geballe at Bell Labs
 had discovered 
SmB$_{6}$\cite{smb6}, 
in which the magnetic Sm ions avoid ordering by developing 
a narrow-gap insulator, now called a ``Kondo insulator''.  
In 1975, an ETH Zurich-Bell Labs collaboration discovered the first heavy
fermion metal CeAl$_{3}$\cite{ott75}. 
The amazing thing about these two materials,
is that both display the same sort of Kondo resistance scattering
at high temperatures, but at low temperatures 
two materials respond differently - with the resistivity
sky-rocketing in SmB$_{6}$, but collapsing into a coherent 
low temperature 
metal in CeAl$_{3}$. Three years later, Frank
Steglich discovered  the first heavy fermion superconductor
CeCu$_{2}$Si$_{2}$\cite{steglich}, though it took a number of years for
the community to change their mind-set and accept this pioneering discovery. 
\fg{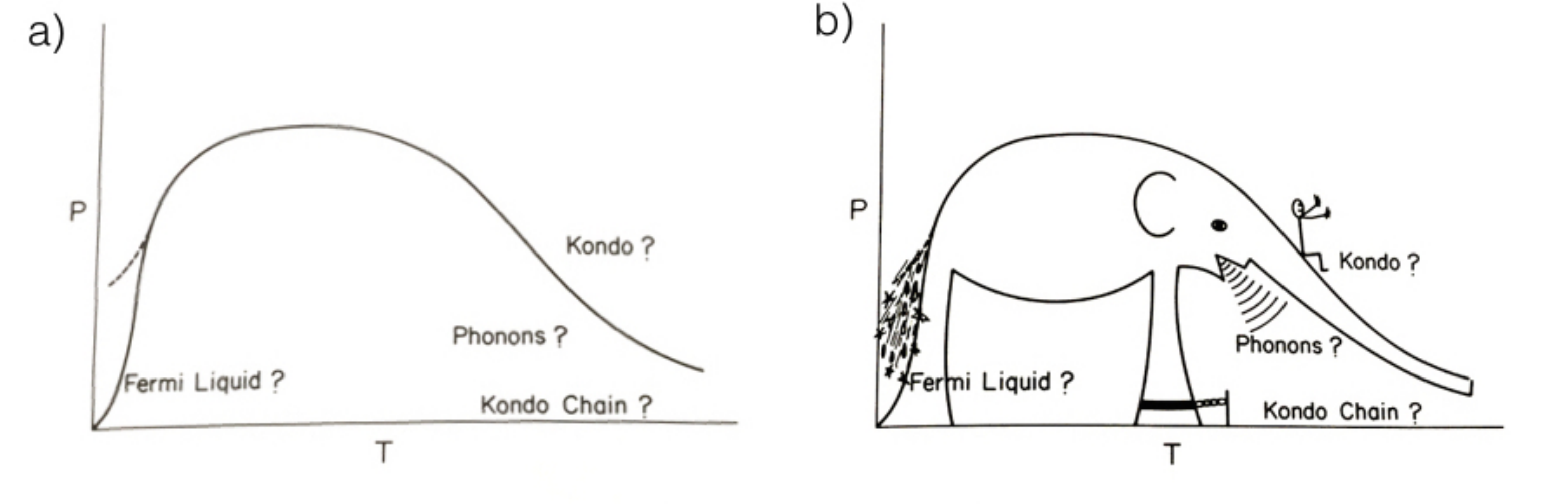}{elephant}{Sketches from Phil Anderson's ``Epilogue''
from the 1976 Rochester Conference on Mixed
Valence\cite{epilogue}. (a) Resistivity stereotypical of systems such as
CeAl$_{3}$. (b) Elephantine version of Fig a).  (Art work by
PWA). Reproduced from Valence Instabilities and Narrow Band
Phenomenon, P. W. Anderson, editor Ron Parks, p 389-396 (1977) with
permission from the author. }

Despite this rapid progress on the experimental front, theoretical
progress was flummoxed by the difficulty of making the 
transition to the dense  ``Kondo lattice'' problem, lacking both the
conceptual and mathematical framework. 
Phil's  input, particularly his summary talks at the 1976 Rochester
and 1980 Santa Barbara meetings on mixed valence had a profound
impact. 

Ron Parks and Chandra Varma organized the first conference on mixed
valence at the University of Rochester in November 1976 and invited 
Phil to give the summary. 
As part of this summary, Phil roasted the theory community 
by sketching the resistivity of heavy fermion metals (see Fig
\ref{elephant}a) in the guise of an elephant.  Recalling an
Indian parable about an elephant and seven blind men, one who
pulls its tail and says its a rope, the other who says its leg is tree
and so on, Phil introduced his elephantine sketch of Kondo lattice 
resistivity, with Jun Kondo sliding down the elephant's trunk and a Fermi
liquid coming out of its behind! (See Fig. \ref{elephant}b ).
\footnote{Did this sketch reveal 
Phil's subconscious discomfort with Landau Fermi liquid theory ?}
The main point of the figure however, was 
to urge the theory community to unify its understanding 
of these diverse phenomena. 

In the transcript describing the Kondo elephant,  Phil remarks 
\begin{quotation}
{\sl ``Now we come to the heart and core of the elephant, the part which
nobody has really done, which was first mentioned at least as a
serious problem here in this conference, namely the Kondo lattice,
which Seb Doniach has made a start on. What you really have here is a
lattice full of these objects that fluctuate back and forth from one
valence to another. There are the phonons, there is the fact that the
electrons fluctuate by tossing electrons into the d level on the next
site which can then go down into the f levels on yet another site. So
the things which toss the valence back and forth are definitely
coupled between one site and another. The net result of doing this is
something that most of the experiments have to tell us about : that this
probably renormalizes to a very heavy Fermi liquid theory with some
kind of strong antiferromagnetic prejudice in that the f-like objects
in the Fermi liquid somehow lost all of their desire to be
magnetic and don't very easily order anymore. This is an extremely
hard problem, it's a problem in the same category of problems which
are failing to be done in field theory these days.''}
\end{quotation}
In this brief paragraph, Phil has laid out his view of 
the physical framework needed to understand heavy fermion
materials. Despite his cartoon, he did 
emphasize that the low temperature
ground-state would be a renormalized heavy Fermi liquid. 
He also notes the parallel between strongly correlated materials and
the challenges of field theory, a parallel that would inspire
many younger physicists in the decades to come. 
Later in the talk,
Phil discusses the possibility of further instabilities in the
Fermi liquid, and expresses the view that these will be more than just 
antiferromagnets: 
\begin{quotation}
{\sl ``Once you get down to this Fermi liquid, it seems that there is a
serious question of what then happens? What does the resulting heavy
Fermi liquid do with itself, what further transformation might it
undergo? There are several possibilities. $\dots $ There is no reason
at all why it shouldn't localize and maybe there are cases where it
localizes. Kasuya gave an argument for one of then. A second
possibility a whole series of experiments seem to indicate is that
some phase transition takes place in many cases. the question is: what
is the nature of these phase transitions? I for one am not ready to
accept the idea that they are all simple magnetic phase transitions
$\dots $. Maybe there is some kind of d to f excitonic phase
transition that either does or does not leave some Fermi surface
behind. Maybe there's a density wave. What else? '' }
\end{quotation}
Curiously though, reflecting the continued mind-set of the community
Phil does not mention the possibility of superconductivity, reflecting
the fact that Steglich's 1979 work was not yet widely accepted
\footnote{
Indeed, although Phil didn't know it, superconductivity had been seen
at Bell Labs three years earlier in the heavy fermion material
UBe$_{13}$\cite{bucher}, but mis-interpreted as an artifact of uranium filaments.}

In 1980, Walter Kohn, Brian Maple and Werner Hanke at 
the Institute for Theoretical Physics, Santa Barbara (now the Kavli
Institute for Theoretical Physics) 
organized a six month workshop on valence fluctuations, culminating in
a conference in January 1981.  In the summary of the conference, 
Phil continued on the theme of the link between field theory and
strongly correlated electrons, introducing for the first time, the
seminal idea that a {\sl large $N$ expansion}, akin to that used in
particle physics,  might be useful.

Phil magnetic life already had two links with the idea of a large
$N$ expansion. Of course, his early work on spin-wave theory was based on
a $1/S$ expansion, but more recently, 
his work with Sam Edwards on the infinite
range spin glass had involved 
the replica trick, replacing the
disorder-averaged Free energy $\overline{-T\ln Z}$ 
with the $N\rightarrow 0$
limit of the disorder-averaged partition function 
average of $N$ replicas, 
\begin{equation}\label{}
\overline{\ln Z} = \lim_{N\rightarrow 0}{\overline{\left(Z^{N}-1 \right)/N}}.
\end{equation}
To take the $N\rightarrow 0$ limit 
requires that one first solve the problem at 
large $N$ to extrapolate back to zero.

But the context of heavy fermions Phil noticed that there was already
a large finite $N$ to expand in. 
Rare earth atoms are strongly
spin-orbit coupled, and so, ignoring crystal field effects, they have
a large spin degeneracy $N=2j+1$, where $j=5/2$ or $7/2$ for
individual f-electrons. Phil realized that the 
parameter $1/N$ could act as an effective small parameter for
resuming many body effects:
\begin{quotation}
{\sl ``  The most important one, $\dots $
is the importance of what you might call the
large $N$ limit; it was only at this conference that I, at least,
realized that we have been going through a case of parallel evolution
with non-Abelian gauge theory. This really has great resemblances to
what one does in the intermediate valence problem, and it is
interesting that the gauge theorists have found that their best
controlled approximations are in a limit which they call large $N$ -
which is large order of the group, large degeneracy of the particles,
and in our case that has to do with large values of the degeneracies
of the states. This is the number that Ramakrishnan called
$n_{\lambda}$. I'm going to talk later about how many different kinds
of roles that plays.''}
\end{quotation}
Later in the same article, Phil expands on this idea and how it can
be used for scaling.  He makes two key observations:
\begin{itemize}
\item That valence fluctuations are $N$-fold enhanced by the large
orbital degeneracy of f-electrons. 
\item That intersite interactions are reduced by a factor of $1/N$
relative to onsite interactions. 
\end{itemize}

Summarizing a full page of discussion, Phil writes
\begin{quotation}
{\sl ``So we find again and again that we are gaining from this
degeneracy factor and it may make the problem a lot simpler than
such apparently easier problems like the Kondo problem.''}
\end{quotation}

Phil's new proposal 
had an electrifying effect on the fledgling strongly correlated
electron theory community, 
for it undid the log-jam, providing for the first time, a controllable
expansion parameter for dealing with the mixed valent and Kondo
lattices.  
In the immediate future, A wide range of large $N$ treatments of the
Anderson model followed, including work by Ramakrishnan and Sur\cite{ramalargeN,ramakrishnansur}, Gunnarson and
Schonhammer\cite{gunnarson} and by  Zhang and Lee\cite{zhanglee}.  
Phil's observations also inspired a
search for a more field theoretic way to formulate the Kondo and mixed
valence problems, leading to the pioneering work by Nicolas Read and
Dennis Newns\cite{read83_1,read83_v2} on 
the large $N$ Kondo model and my own {\sl slave boson} approach\cite{me1983} to mixed
valence developed under Phil's generous tutelage, in which the
Gutzwiller projected f-electron operator is factorized in terms of an
Abrikosov pseudo-fermion and a slave boson operator $X_{\sigma
0}= f\dg_{\sigma }b$.  With this device, one could  see for the first
time, that the no-double occupancy constraint gives rise to 
locally conserved charges (here $n_{b}+n_{f}=Q$) and corresponding
gauge fields.

\section{Superconductivity and Magnetism come together}\label{ra_sec1}

\fg{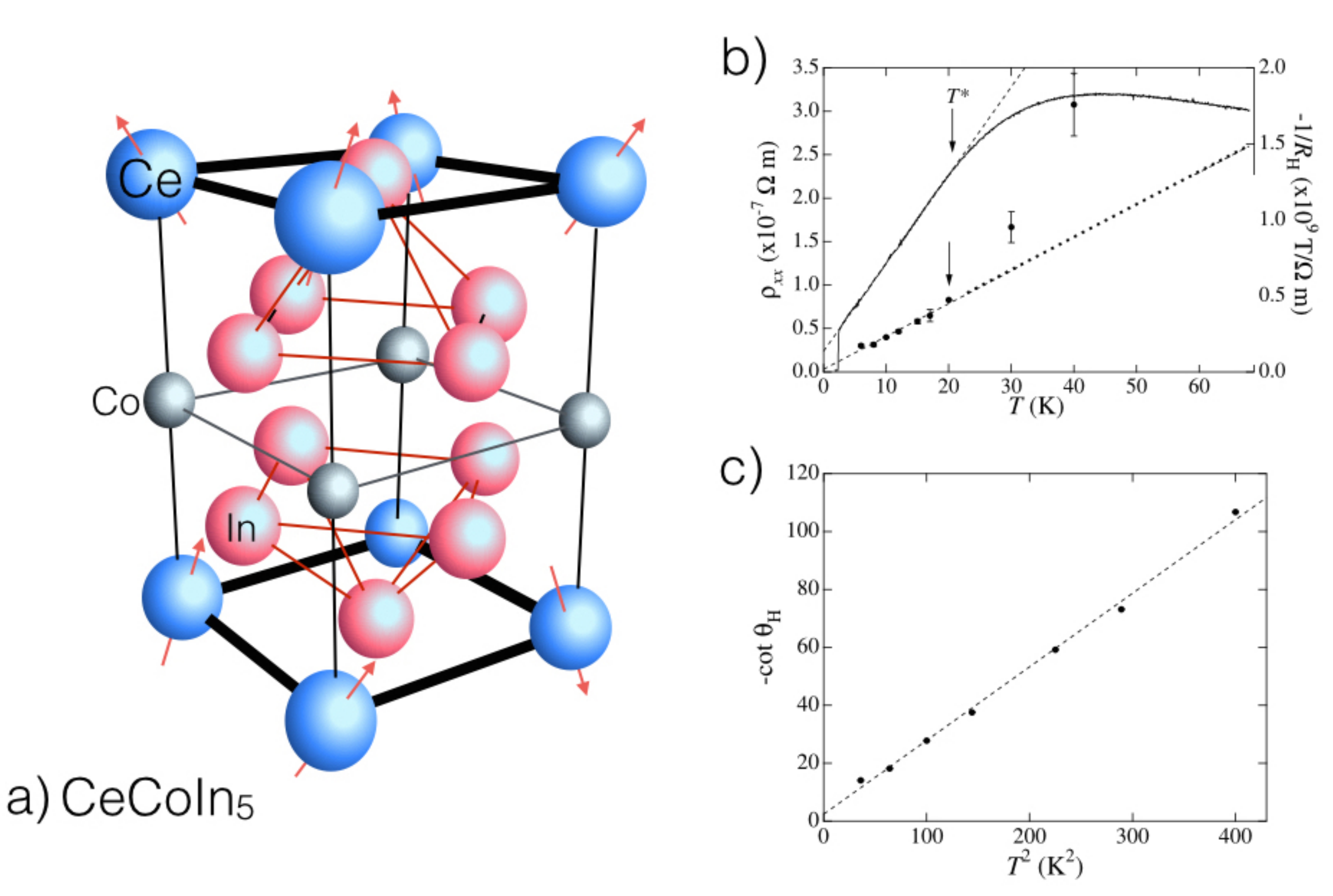}{115}{The heavy fermion superconductor CeCoIn$_{5}$
showing a) the structure of this layered compound (b) the resistivity
and inverse Hall constant, which are both linearly proportional to the
temperature.  The linear temperature dependence of the resistivity
indicates a linear temperature dependence of the electron scattering
rate.  The temperature dependence of the Hall constant indicates that 
the Hall transport relaxation rate and the linear transport
relaxataion rate are not equal after \cite{Nakajima:2004wy}. 
 c) the cotangent of the Hall angle, showing the $T^{2}$
dependence of the Hall scattering rate after \cite{Nakajima:2004wy}. 
}

The theoretical perspective of condensed matter physics has 
dramatically transformed over the period of Phil's research. 
Fifty years ago, magnetism and superconductivity were
regarded as mutually exclusive forms of order.
Yet, gradually, starting in the 1970s, 
the discovery of new kinds of pair condensate, 
of superfluid He-3\cite{Osheroff1972}, of heavy fermion
\cite{steglich}, 
organic\cite{jerome},  high temperature cuprate 
\cite{bednorz} and  iron-based superconductors\cite{hosono}, 
has indicated a more intimate connection between magnetism and
superconductivity. 

 Phil's  ideas have evolved during this same period, and 
as they have done so, 
they have often transformed our scientific consensus. 
Phil started in the 1950s accounting for the
stability of antferromagnetic order against quantum fluctuations, 
at the time itself controversial.
Through a journey via the Kondo effect, spin liquids and spin glasses
he was led to consider states of matter in which conventional magnetism
is absent and the magnetic degrees of freedom drive
new kinds of electronic ground-states, especially superconductivity.

I'd like to selectively mention three exciting areas of evolving and 
currently unresolved controversy connected with Phil's  ideas. 

\begin{enumerate}

\item \underline{Mott versus Landau.} 
From the very outset,  Phil
he has emphasized the importance of the {\sl Mott
mechanism}, namely the exclusion of  double occupancy of atomic
orbitals. 
One of the questions he has emphasized,
is whether the Landau quasiparticle description of electrons 
can survive the imposition of these severe constraints,   
suggesting instead that 
new kinds of metallic ground-states must inevitably develop 
in which the excitations have zero overlap with non-interacting electrons,
and thus can not be regarded as Landau quasiparticles. 
Central to Phil's arguments, is the idea that the 
of electrons to highly constrained electron fluids leads to a
many-body X-ray catastrophe that leads to the inevitable demise of
the Landau quasiparticle to form {\sl strange
metals}\cite{philstrange1,philstrange2}, and sometimes, 
{\sl hidden Fermi liquids}\cite{phil_hidden,phil_hidden2}, 
which resemble the Landau Fermi liquid
thermodynamically, but without overlap with the original electron
fields. 

\item \underline{Quantum Criticality versus Strange Metals}
There are now
many example of metals which exhibit highly unusual transport and
thermodynamic properties which defy a Landau Fermi liquid
description, such as the optimally doped normal state of the cuprate
superconductors\cite{ong}, MnSi under pressure 
and various heavy fermion materials\cite{lonzarichmnsi}, such as
CeCoIn$_{5}$\cite{Tanatar,Nakajima:2004wy} (see Fig. \ref{115})
and YbAlB$_{4}$\cite{satoru}
which each exhibit unusual linear or power-law
temperature dependencies in the resistivity.
One of the key discussions about these materials is whether such non-Fermi
liquid behavior is generated by the vicinity to a {\sl quantum
critical point}, or whether, as Phil believes, 
the unusual metallic behavior is 
related to a new kind {\sl strange metal phase}\cite{phil_hidden2}.
The recent discovery of a pressure-independent
anomalous metal phase in YbAlB$_{4}$ may be an example of such a strange
metal phase\cite{satoru}. 

\item \underline{Fabric versus Glue.}
The conventional view of 
unconventional superconductors argues that they should be regarded as
magnetic analogs of phonon-mediated superconductors, in which the
soft magnetic fluctuations provide the 
pairing {\sl glue}. Phil has argued\cite{PW:2007ei} for a different picture, in which
pre-formed, resonating valence bonds, on doping, provide the
underlying {\sl fabric} for a pair condensate.  These opposing 
ideas continue to be lively debated in the context of high-temperature 
cuprate superconductors.   Another place they may be important, is in heavy
fermion superconductors, where the Kondo effect can play the same role
as doping, forcing valence bonds out into the conduction sea to form pairs.

\end{enumerate}

In his 2006 paper, {\sl ``The ‘strange metal’ is a projected Fermi
liquid with edge singularities''}\cite{philstrange2}, Phil summarizes his point of view, writing
\begin{quotation}
{``\sl This ‘strange metal’ phase continues to be of much
theoretical interest. Here we show it is a consequence of projecting
the doubly occupied amplitudes out of a conventional Fermi- sea
wavefunction (Gutzwiller projection), requiring no exotica such as a
mysterious quantum critical point. Exploiting a formal similarity with
the classic problem of Fermi-edge singularities in the X-ray spectra
of metals, we find a Fermi-liquid-like excitation spectrum, but the
excitations are asymmetric between electrons and holes, show anomalous
forward scattering and the renormalization constant Z = 0.''}
\end{quotation}

One of the most fascinating, and still unsolved aspects of these
discussions above concerns the apparent 
development of two transport lifetimes in the electronic
conductivity\cite{ong,Anderson:1991wz}:
a transport scattering lifetime, inversely
proportional to temperature $\tau_{tr}^{-1}\propto k_{B}T$ and a Hall
scattering time, inversely proportional the square of the temperature
$\tau_{H}^{-1}\propto T^{2}$.  
In a modified Drude formalism, the linear and Hall conductivities are
given by 
\begin{eqnarray}\label{l}
\sigma_{xx} &=& 
\frac{ne^{2}}{m} \tau_{tr}
\cr \sigma_{xy}&=& \frac{ne^{2}}{m} \tau_{tr} (\omega_{c}\tau_{H}),
\end{eqnarray}
giving rise to a resistivity $\rho_{xx}\propto \tau_{tr}^{-1}\sim T$
and a  Hall angle which satisfies  $\cot \theta_{H}=
\sigma_{xx}/\sigma_{xy}\propto \tau_{H}^{-1}\propto T^{2}$. 
There are now three separate classes of material where this behavior
has been seen: the cuprate metals\cite{ong}, 
the 115 heavy fermion superconductors CeCoIn$_{5}$\cite{Nakajima:2004wy}
and  electrons fluids at two dimensional oxide interfaces
(SrTiO$_{3}$/RTiO$_{3}$ (R=Gd,Sm))\cite{Mikheev:2015cb}. 
The remarkable aspect of these metals,
is that the two relaxation times enter {\sl
multiplicatively } into their Hall conductivity, 
$\sigma_{xy}\propto \tau_{tr}\tau_{H}$. 
Since since $\sigma_{xy}$ is a zero momentum probe of the current
fluctuations at the Fermi
surface, this suggests 
that electrons are subject to two separate relaxation times
at the very same point on the Fermi surface, linked by 
the current operator. 
Phil's ideas on this subject \cite{Anderson:1991wz}
have inspired a range of new theories
\cite{twotimes,Hussey:2008tw,Blake:2015gc}, but we still await a final
understanding. 

Like many in our community, I've often marveled at Phil Anderson's
ability to radically transform his viewpoints in response to new data and new
insights.  I've asked him what it would be like if he ever met his
younger self for a physics discussion, and he agrees that he'd
probably have quite a forceful disagreement on topics he
originally pioneered and on which he now has a new perspective.
Perhaps Tom Stoppard will write a play on this someday. 

Phil, here's to the continuing success and inspiration of your magnetic ideas!


\section*{Acknowledgments}
In writing this article I have benefited from conversations with
Natan Andrei,  Premala Chandra and Don Hamann. 
This work was supported by NSF grant DMR-1309929.

\end{document}

%% file: defs.tex
\def\3he{{$^3${\rm He}}}

\def\slD{\raise.15ex\hbox{$/$}\kern-.57em\hbox{$D$}}
\def\dsl{\raise.15ex\hbox{$/$}\kern-.57em\hbox{$\Delta$}}
\def\slp{{\raise.15ex\hbox{$/$}\kern-.57em\hbox{$\partial$}}}
\def\nsl{\raise.15ex\hbox{$/$}\kern-.57em\hbox{$\nabla$}}
\def\sla{\raise.15ex\hbox{$/$}\kern-.57em\hbox{$\rightarrow$}}
\def\slla{\raise.15ex\hbox{$/$}\kern-.57em\hbox{$\lambda$}}
\def\gtwid{\raise.3ex\hbox{$>$\kern-.75em\lower1ex\hbox{$\sim$}}}
\def\ltwid{\raise.3ex\hbox{$<$\kern-.75em\lower1ex\hbox{$\sim$}}}

\def\12{{1\over2}}

\def\part{\partial}

\def\bk{{\bf k}}

\def\bR{{\bf R}}

\def\bethlogo{\vbox{\bf \line{\hrulefill} 
    \kern-.5\baselineskip 
    \line{\hrulefill\phantom{ ELIZABETH A. MASON }\hrulefill} 
    \kern-.5\baselineskip 
    \line{\hrulefill\hbox{ ELIZABETH A. MASON }\hrulefill} 
    \kern-.5\baselineskip 
    \line{\hrulefill\phantom{ 1411 Chino Street }\hrulefill} 
    \kern-.5\baselineskip 
    \line{\hrulefill\hbox{ 1411 Chino Street }\hrulefill} 
    \kern-.5\baselineskip 
    \line{\hrulefill\phantom{ Santa Barbara, CA 93101 }\hrulefill} 
    \kern-.5\baselineskip 
    \line{\hrulefill\hbox{ Santa Barbara, CA 93101 }\hrulefill}
    \kern-.5\baselineskip 
    \line{\hrulefill\phantom{ (805) 962-2739 }\hrulefill} 
    \kern-.5\baselineskip 
    \line{\hrulefill\hbox{ (805) 962-2739 }\hrulefill}}}
\def\lisalogo{\vbox{\bf \line{\hrulefill} 
    \kern-.5\baselineskip 
    \line{\hrulefill\phantom{ LISA R. GOODFRIEND }\hrulefill} 
    \kern-.5\baselineskip 
    \line{\hrulefill\hbox{ LISA R. GOODFRIEND }\hrulefill} 
    \kern-.5\baselineskip 
    \line{\hrulefill\phantom{ 6646 Pasado }\hrulefill} 
    \kern-.5\baselineskip 
    \line{\hrulefill\hbox{ 6646 Pasado }\hrulefill} 
    \kern-.5\baselineskip 
    \line{\hrulefill\phantom{ Santa Barbara, CA 93108 }\hrulefill} 
    \kern-.5\baselineskip 
    \line{\hrulefill\hbox{ Santa Barbara, CA 93108 }\hrulefill}
    \kern-.5\baselineskip 
    \line{\hrulefill\phantom{ (805) 962-2739 }\hrulefill} 
    \kern-.5\baselineskip 
    \line{\hrulefill\hbox{ (805) 962-2739 }\hrulefill}}}

\def\low#1{\lower.5ex\hbox{${}_#1$}}
\def\ltwid{\raise.3ex\hbox{$<$\kern-.75em\lower1ex\hbox{$\sim$}}}

\def\psl{\raise.15ex\hbox{$/$}\kern-.57em\hbox{$\partial$}}
\def\partt{\raise.15ex\hbox{$\widetilde$}{\kern-.37em\hbox{$\partial$}}}
\def\parts{\raise.15ex\hbox{$/$}{\kern-.6em\hbox{$\partial$}}}
\def\nablas{\raise.15ex\hbox{$/$}{\kern-.6em\hbox{$\nabla$}}}
\def\oprod{\hbox{$\rm O$}{\kern -0.8em\hbox{$\Pi$}}}
\def\partw#1{\raise.15ex\hbox{$/$}{\kern-.6em\hbox{${#1}$}}}

\def\gtappr{{{\lower4pt\hbox{$>$} } \atop \widetilde{ \ \ \ }}}
\def\ltappr{{{\lower4pt\hbox{$<$} } \atop \widetilde{ \ \ \ }}}

\def\topppageno1{\global\footline={\hfil}\global\headline
={\ifnum\pageno<\firstpageno{\hfil}\else{\hss\twelverm --\ \folio
\ --\hss}\fi}}

\def\toppageno2{\global\footline={\hfil}\global\headline
={\ifnum\pageno<\firstpageno{\hfil}\else{\rightline{\hfill\hfill
\twelverm \ \folio
\ \hss}}\fi}}

\def\ltdash{\raise-1.8pt\hbox{$\scriptscriptstyle |$}}

\newlength{\upit}\upit=0.1truein

\newlength{\bxwidth}\bxwidth=1.5 truein

\newcommand{\dg}{^{\dagger }}

\newcommand{\up}{\uparrow}
\newcommand{\dw}{\downarrow}

\newcommand{\pmat}[1]{\begin{pmatrix} #1 \end{pmatrix}}

\newlength{\figwidth}
\newlength{\shift}
\newlength{\fight}
\newcommand{\fg}[3]
{
\begin{figure}[ht]
\vspace*{-0cm}
\[
\includegraphics[width=\fight]{#1}
\]
\vskip -0.2cm
\caption{\label{#2}
\small#3
}
\end{figure}}

\newcommand \bea {\begin{eqnarray} }
\newcommand \eea {\end{eqnarray}}

\newsavebox{\fmbox}

%% file: defs2.tex
\def\dg{{^
{\dag}}}

\def\1{{\bf 1}}
\def\2{{\bf 2}}

\def\ell{{\it l } {\rm n}}

\def\cx2{\sqrt{c^2_x+c^2_y}}

\def\gkk{\gamma _{\vec k}}
\def\gk2{\gkk ^2}
\def\dw{\downarrow}
\def\up{\uparrow}
\def\gtappr{{{\lower4pt\hbox{$>$} } \atop \widetilde{ \ \ \ }}}
\def\ltappr{{{\lower4pt\hbox{$<$} } \atop \widetilde{ \ \ \ }}}

\def\pbar{{\partial\kern-1.2ex\raise0.25ex\hbox{/}}}

\def\up{\uparrow}
\def\dw{\downarrow}

\def\dg{{^{\dag}}}

\def\1{{\bf 1}}
\def\2{{\bf 2}}

\def\ell{{\it l } {\rm n}}

\def\cx2{\sqrt{c^2_x+c^2_y}}

\def\gkk{\gamma _{\vec k}}
\def\gk2{\gkk ^2}
\def\gtappr{{{\lower4pt\hbox{$>$} } \atop \widetilde{ \ \ \ }}}
\def\ltappr{{{\lower4pt\hbox{$<$} } \atop \widetilde{ \ \ \ }}}
\def\bq{{\bf q}}

\def\thickra{\hbox{\raise0.2pt\hbox{{$\bf >\mkern-13mu>\mkern-13mu>$}}}}
\def\thickrarrow{\hbox{\raise0.28pt\hbox{{$\bf >\mkern-13mu>\mkern-13mu>$}}}}

%% file: pwaml_arXivX.bbl
\begin{thebibliography}{75}%
\makeatletter
\providecommand \@ifxundefined [1]{%
 \@ifx{#1\undefined}
}%
\providecommand \@ifnum [1]{%
 \ifnum #1\expandafter \@firstoftwo
 \else \expandafter \@secondoftwo
 \fi
}%
\providecommand \@ifx [1]{%
 \ifx #1\expandafter \@firstoftwo
 \else \expandafter \@secondoftwo
 \fi
}%
\providecommand \natexlab [1]{#1}%
\providecommand \enquote  [1]{``#1''}%
\providecommand \bibnamefont  [1]{#1}%
\providecommand \bibfnamefont [1]{#1}%
\providecommand \citenamefont [1]{#1}%
\providecommand \href@noop [0]{\@secondoftwo}%
\providecommand \href [0]{\begingroup \@sanitize@url \@href}%
\providecommand \@href[1]{\@@startlink{#1}\@@href}%
\providecommand \@@href[1]{\endgroup#1\@@endlink}%
\providecommand \@sanitize@url [0]{\catcode `\\12\catcode `\$12\catcode
  `\&12\catcode `\#12\catcode `\^12\catcode `\_12\catcode `\%12\relax}%
\providecommand \@@startlink[1]{}%
\providecommand \@@endlink[0]{}%
\providecommand \url  [0]{\begingroup\@sanitize@url \@url }%
\providecommand \@url [1]{\endgroup\@href {#1}{\urlprefix }}%
\providecommand \urlprefix  [0]{URL }%
\providecommand \Eprint [0]{\href }%
\providecommand \doibase [0]{http://dx.doi.org/}%
\providecommand \selectlanguage [0]{\@gobble}%
\providecommand \bibinfo  [0]{\@secondoftwo}%
\providecommand \bibfield  [0]{\@secondoftwo}%
\providecommand \translation [1]{[#1]}%
\providecommand \BibitemOpen [0]{}%
\providecommand \bibitemStop [0]{}%
\providecommand \bibitemNoStop [0]{.\EOS\space}%
\providecommand \EOS [0]{\spacefactor3000\relax}%
\providecommand \BibitemShut  [1]{\csname bibitem#1\endcsname}%
\let\auto@bib@innerbib\@empty
\bibitem [{\citenamefont {Anderson}(1984)}]{andersonconcepts}%
  \BibitemOpen
  \bibfield  {author} {\bibinfo {author} {\bibfnamefont {P.~W.}\ \bibnamefont
  {Anderson}},\ }\href@noop {} {\emph {\bibinfo {title} {{\sl Basic Notions of
  Condensed Matter Physics}}}}\ (\bibinfo  {publisher} {Benjamin Cummings},\
  \bibinfo {year} {1984})\BibitemShut {NoStop}%
\bibitem [{\citenamefont {N\'eel}(1932)}]{Neel}%
  \BibitemOpen
  \bibfield  {author} {\bibinfo {author} {\bibfnamefont {Louis}\ \bibnamefont
  {N\'eel}},\ }\bibfield  {title} {\enquote {\bibinfo {title} {{\sl Influence
  des fluctuations du champ mol\' eculaires sur les propri\' et\' es
  magn\' etiques des
  corps}},}\ }\href {https://tel.archives-ouvertes.fr/tel-00278201v1}
  {\bibfield  {journal} {\bibinfo  {journal} {Ann. de Physique}\ }\textbf
  {\bibinfo {volume} {18}},\ \bibinfo {pages} {5--105} (\bibinfo {year}
  {1932})}\BibitemShut {NoStop}%
\bibitem [{\citenamefont {Landau}(1933)}]{Landau}%
  \BibitemOpen
  \bibfield  {author} {\bibinfo {author} {\bibfnamefont {Lev~D.}\ \bibnamefont
  {Landau}},\ }\bibfield  {title} {\enquote {\bibinfo {title} {{\sl A possible
  explanation of the field dependence of the suseptibility at low
  temperatures}},}\ }\href@noop {} {\bibfield  {journal} {\bibinfo  {journal}
  {Phys. Z. Sowjet}\ }\textbf {\bibinfo {volume} {4}},\ \bibinfo {pages} {675}
  (\bibinfo {year} {1933})}\BibitemShut {NoStop}%
\bibitem [{\citenamefont {Bethe}(1931)}]{bethe}%
  \BibitemOpen
  \bibfield  {author} {\bibinfo {author} {\bibfnamefont {H.}~\bibnamefont
  {Bethe}},\ }\bibfield  {title} {\enquote {\bibinfo {title} {{\sl Zur Theorie
  der Metalle: I. Eigenwerte und Eigenfunktionen der linearer Atomkerte, (On
  the Theory of metals: I. Eigenvalues and Eigenfunctions of the linear atom
  chain)}},}\ }\href {http://dx.doi.org/10.1007/BF01341708} {\bibfield
  {journal} {\bibinfo  {journal} {Zeitschrift fur Physik}\ }\textbf {\bibinfo
  {volume} {71}},\ \bibinfo {pages} {205--226} (\bibinfo {year}
  {1931})}\BibitemShut {NoStop}%
\bibitem [{\citenamefont {Pomeranchuk}(1941)}]{pomeranchuk}%
  \BibitemOpen
  \bibfield  {author} {\bibinfo {author} {\bibfnamefont {I.}~\bibnamefont
  {Pomeranchuk}},\ }\bibfield  {title} {\enquote {\bibinfo {title} {{\sl
  Thermal conductivity of paramagnetic insulators at low temperatures}},}\
  }\href@noop {} {\bibfield  {journal} {\bibinfo  {journal} {Zh. Eksp. Teor.
  Fiz}\ }\textbf {\bibinfo {volume} {11}},\ \bibinfo {pages} {226} (\bibinfo
  {year} {1941})}\BibitemShut {NoStop}%
\bibitem [{\citenamefont {Anderson}(1952)}]{1952}%
  \BibitemOpen
  \bibfield  {author} {\bibinfo {author} {\bibfnamefont {P.~W.}\ \bibnamefont
  {Anderson}},\ }\bibfield  {title} {\enquote {\bibinfo {title} {{\sl An
  Approximate Quantum Theory of the Antiferromagnetic Ground State}},}\ }\href
  {http://link.aps.org/doi/10.1103/PhysRev.86.694} {\bibfield  {journal}
  {\bibinfo  {journal} {Phys. Rev.}\ }\textbf {\bibinfo {volume} {86}},\
  \bibinfo {pages} {694--701} (\bibinfo {year} {1952})}\BibitemShut {NoStop}%
\bibitem [{\citenamefont {Shull}\ and\ \citenamefont
  {Smart}(1949)}]{shullsmart}%
  \BibitemOpen
  \bibfield  {author} {\bibinfo {author} {\bibfnamefont {C.~G.}\ \bibnamefont
  {Shull}}\ and\ \bibinfo {author} {\bibfnamefont {J.~Samuel}\ \bibnamefont
  {Smart}},\ }\bibfield  {title} {\enquote {\bibinfo {title} {{\sl Detection of
  Antiferromagnetism by Neutron Diffraction}},}\ }\href
  {http://link.aps.org/doi/10.1103/PhysRev.76.1256.2} {\bibfield  {journal}
  {\bibinfo  {journal} {Phys. Rev.}\ }\textbf {\bibinfo {volume} {76}},\
  \bibinfo {pages} {1256--1257} (\bibinfo {year} {1949})}\BibitemShut {NoStop}%
\bibitem [{\citenamefont {Klein}\ and\ \citenamefont
  {Smith}(1950)}]{kleinsmith}%
  \BibitemOpen
  \bibfield  {author} {\bibinfo {author} {\bibfnamefont {M.~J}\ \bibnamefont
  {Klein}}\ and\ \bibinfo {author} {\bibfnamefont {R.S}\ \bibnamefont
  {Smith}},\ }\bibfield  {title} {\enquote {\bibinfo {title} {{\sl A Note on
  the Classical Spin-Wave Theory of Heller and Kramers}},}\ }\href
  {http://link.aps.org/doi/10.1103/PhysRev.80.1111} {\bibfield  {journal}
  {\bibinfo  {journal} {Phys. Rev.}\ }\textbf {\bibinfo {volume} {80}},\
  \bibinfo {pages} {1111} (\bibinfo {year} {1950})}\BibitemShut {NoStop}%
\bibitem [{\citenamefont {Kubo}(1952)}]{kubo}%
  \BibitemOpen
  \bibfield  {author} {\bibinfo {author} {\bibfnamefont {Ryogo}\ \bibnamefont
  {Kubo}},\ }\bibfield  {title} {\enquote {\bibinfo {title} {{\sl The Spin-Wave
  Theory of Antiferromagnetics}},}\ }\href
  {http://link.aps.org/doi/10.1103/PhysRev.87.568} {\bibfield  {journal}
  {\bibinfo  {journal} {Phys. Rev.}\ }\textbf {\bibinfo {volume} {87}},\
  \bibinfo {pages} {568--580} (\bibinfo {year} {1952})}\BibitemShut {NoStop}%
\bibitem [{\citenamefont {Fazekas}\ and\ \citenamefont
  {Anderson}(1974)}]{fazekas}%
  \BibitemOpen
  \bibfield  {author} {\bibinfo {author} {\bibfnamefont {P.}~\bibnamefont
  {Fazekas}}\ and\ \bibinfo {author} {\bibfnamefont {P.~W.}\ \bibnamefont
  {Anderson}},\ }\bibfield  {title} {\enquote {\bibinfo {title} {{\sl On the
  ground state properties of the anisotropic triangular antiferromagnet}},}\
  }\href {http://www.tandfonline.com/doi/abs/10.1080/14786439808206568}
  {\bibfield  {journal} {\bibinfo  {journal} {Philos. Mag.}\ }\textbf {\bibinfo
  {volume} {30}},\ \bibinfo {pages} {423--440} (\bibinfo {year}
  {1974})}\BibitemShut {NoStop}%
\bibitem [{\citenamefont {Haldane}(1982)}]{haldane82}%
  \BibitemOpen
  \bibfield  {author} {\bibinfo {author} {\bibfnamefont {F.~D.~M.}\
  \bibnamefont {Haldane}},\ }\bibfield  {title} {\enquote {\bibinfo {title}
  {{\sl Continuum dynamics of the 1-D Heisenberg antiferromagnet:
  Identification with the O(3) nonlinear sigma model}},}\ }\href
  {http://www.sciencedirect.com/science/article/pii/037596018390631X}
  {\bibfield  {journal} {\bibinfo  {journal} {Physics Letters A}\ }\textbf
  {\bibinfo {volume} {93}},\ \bibinfo {pages} {464--468} (\bibinfo {year}
  {1982})}\BibitemShut {NoStop}%
\bibitem [{\citenamefont {Anderson}(1958)}]{pwarpa}%
  \BibitemOpen
  \bibfield  {author} {\bibinfo {author} {\bibfnamefont {P.~W.}\ \bibnamefont
  {Anderson}},\ }\bibfield  {title} {\enquote {\bibinfo {title} {{\sl
  Random-Phase Approximation in the Theory of Superconductivity}},}\ }\href
  {http://link.aps.org/doi/10.1103/PhysRev.112.1900} {\bibfield  {journal}
  {\bibinfo  {journal} {Phys. Rev.}\ }\textbf {\bibinfo {volume} {112}},\
  \bibinfo {pages} {1900--1916} (\bibinfo {year} {1958})}\BibitemShut {NoStop}%
\bibitem [{\citenamefont {Nambu}(1960)}]{nambu}%
  \BibitemOpen
  \bibfield  {author} {\bibinfo {author} {\bibfnamefont {Yoichiro}\
  \bibnamefont {Nambu}},\ }\bibfield  {title} {\enquote {\bibinfo {title} {{\sl
  Quasi-Particles and Gauge Invariance in the Theory of Superconductivity}},}\
  }\href {http://link.aps.org/doi/10.1103/PhysRev.117.648} {\bibfield
  {journal} {\bibinfo  {journal} {Phys. Rev.}\ }\textbf {\bibinfo {volume}
  {117}},\ \bibinfo {pages} {648--663} (\bibinfo {year} {1960})}\BibitemShut
  {NoStop}%
\bibitem [{\citenamefont {Anderson}(1959)}]{PhysRev.115.2}%
  \BibitemOpen
  \bibfield  {author} {\bibinfo {author} {\bibfnamefont {P.~W.}\ \bibnamefont
  {Anderson}},\ }\bibfield  {title} {\enquote {\bibinfo {title} {{\sl New
  Approach to the Theory of Superexchange Interactions}},}\ }\href
  {http://link.aps.org/doi/10.1103/PhysRev.115.2} {\bibfield  {journal}
  {\bibinfo  {journal} {Phys. Rev.}\ }\textbf {\bibinfo {volume} {115}},\
  \bibinfo {pages} {2--13} (\bibinfo {year} {1959})}\BibitemShut {NoStop}%
\bibitem [{\citenamefont {Anderson}(1968-69{\natexlab{a}})}]{kondoII}%
  \BibitemOpen
  \bibfield  {author} {\bibinfo {author} {\bibfnamefont {P.~W.}\ \bibnamefont
  {Anderson}},\ }\bibfield  {title} {\enquote {\bibinfo {title} {{\sl The Kondo
  Effect. II}},}\ }\href@noop {} {\bibfield  {journal} {\bibinfo  {journal}
  {Comments on Solid State Physics}\ }\textbf {\bibinfo {volume} {I}},\
  \bibinfo {pages} {190} (\bibinfo {year} {1968-69}{\natexlab{a}})}\BibitemShut
  {NoStop}%
\bibitem [{\citenamefont {Anderson}(1994)}]{career}%
  \BibitemOpen
  \bibfield  {author} {\bibinfo {author} {\bibfnamefont {P.~W.}\ \bibnamefont
  {Anderson}},\ }\bibfield  {title} {\enquote {\bibinfo {title} {{\sl The Kondo
  Effect. II}},}\ }in\ \href@noop {} {\emph {\bibinfo {booktitle} {{\sl A
  career in theoretical physics}}}}\ (\bibinfo  {publisher} {World
  Scientific},\ \bibinfo {year} {1994})\ p.\ \bibinfo {pages} {224}\BibitemShut
  {NoStop}%
\bibitem [{\citenamefont {Zener}(1951)}]{zenersd}%
  \BibitemOpen
  \bibfield  {author} {\bibinfo {author} {\bibfnamefont {C.}~\bibnamefont
  {Zener}},\ }\bibfield  {title} {\enquote {\bibinfo {title} {{\sl Interaction
  Between the $d$ Shells in the Transition Metals}},}\ }\href
  {http://link.aps.org/doi/10.1103/PhysRev.81.440} {\bibfield  {journal}
  {\bibinfo  {journal} {Phys. Rev.}\ }\textbf {\bibinfo {volume} {81}},\
  \bibinfo {pages} {440--444} (\bibinfo {year} {1951})}\BibitemShut {NoStop}%
\bibitem [{\citenamefont {Kasuya}(1956)}]{kasuya1956}%
  \BibitemOpen
  \bibfield  {author} {\bibinfo {author} {\bibfnamefont {Tadao}\ \bibnamefont
  {Kasuya}},\ }\bibfield  {title} {\enquote {\bibinfo {title} {{\sl A Theory of
  Metallic Ferro- and Antiferromagnetism on Zener's Model}},}\ }\href
  {http://ptp.oxfordjournals.org/content/16/1/45} {\bibfield  {journal}
  {\bibinfo  {journal} {Progress of Theoretical Physics}\ }\textbf {\bibinfo
  {volume} {16}},\ \bibinfo {pages} {45--57} (\bibinfo {year}
  {1956})}\BibitemShut {NoStop}%
\bibitem [{\citenamefont {Schrieffer}\ and\ \citenamefont
  {Wolff}(1966)}]{swolf}%
  \BibitemOpen
  \bibfield  {author} {\bibinfo {author} {\bibfnamefont {J.~R.}\ \bibnamefont
  {Schrieffer}}\ and\ \bibinfo {author} {\bibfnamefont {P.}~\bibnamefont
  {Wolff}},\ }\bibfield  {title} {\enquote {\bibinfo {title} {{\sl Relation
  between the Anderson and Kondo Hamiltonians{}}},}\ }\href
  {http://link.aps.org/doi/10.1103/PhysRev.149.491} {\bibfield  {journal}
  {\bibinfo  {journal} {Phys. Rev.}\ }\textbf {\bibinfo {volume} {149}},\
  \bibinfo {pages} {491} (\bibinfo {year} {1966})}\BibitemShut {NoStop}%
\bibitem [{\citenamefont {Kondo}(1964)}]{kondo2}%
  \BibitemOpen
  \bibfield  {author} {\bibinfo {author} {\bibfnamefont {J.}~\bibnamefont
  {Kondo}},\ }\bibfield  {title} {\enquote {\bibinfo {title} {{\sl Resistance
  Minimum in Dilute Magnetic Alloys}},}\ }\href
  {http://ptp.oxfordjournals.org/content/32/1/37} {\bibfield  {journal}
  {\bibinfo  {journal} {Prog. Theor. Phys.}\ }\textbf {\bibinfo {volume}
  {32}},\ \bibinfo {pages} {37--49} (\bibinfo {year} {1964})}\BibitemShut
  {NoStop}%
\bibitem [{\citenamefont {de~Haas}\ \emph {et~al.}(1933)\citenamefont
  {de~Haas}, \citenamefont {de~Boer},\ and\ \citenamefont {van~den
  Berg}}]{resistanceminimum}%
  \BibitemOpen
  \bibfield  {author} {\bibinfo {author} {\bibnamefont {de~Haas}}, \bibinfo
  {author} {\bibnamefont {de~Boer}}, \ and\ \bibinfo {author} {\bibfnamefont
  {D.J.}\ \bibnamefont {van~den Berg}},\ }\bibfield  {title} {\enquote
  {\bibinfo {title} {{\sl The electrical resistance of gold, copper and lead at
  low temperatures }},}\ }\href
  {http://www.sciencedirect.com/science/article/pii/S0031891434803102}
  {\bibfield  {journal} {\bibinfo  {journal} {Physica}\ }\textbf {\bibinfo
  {volume} {1}},\ \bibinfo {pages} {1115} (\bibinfo {year} {1933})}\BibitemShut
  {NoStop}%
\bibitem [{\citenamefont {MacDonald}\ and\ \citenamefont
  {Mendelssohn}(1950)}]{resistanceminimum2}%
  \BibitemOpen
  \bibfield  {author} {\bibinfo {author} {\bibfnamefont {D.K.C.}\ \bibnamefont
  {MacDonald}}\ and\ \bibinfo {author} {\bibfnamefont {K.}~\bibnamefont
  {Mendelssohn}},\ }\bibfield  {title} {\enquote {\bibinfo {title} {{\sl
  Resistivity of Pure Metals at Low Temperatures I. The Alkali Metals}},}\
  }\href {\doibase 10.1098/rspa.1950.0117} {\bibfield  {journal} {\bibinfo
  {journal} {Proc. Roy. Soc. London}\ }\textbf {\bibinfo {volume} {202}},\
  \bibinfo {pages} {523} (\bibinfo {year} {1950})}\BibitemShut {NoStop}%
\bibitem [{\citenamefont {Anderson}(1967)}]{xraycat}%
  \BibitemOpen
  \bibfield  {author} {\bibinfo {author} {\bibfnamefont {P.~W.}\ \bibnamefont
  {Anderson}},\ }\bibfield  {title} {\enquote {\bibinfo {title} {{\sl Infrared
  Catastrophe in Fermi Gases with Local Scattering Potentials}},}\ }\href
  {\doibase 10.1103/PhysRevLett.18.1049} {\bibfield  {journal} {\bibinfo
  {journal} {Phys. Rev. Lett.}\ }\textbf {\bibinfo {volume} {18}},\ \bibinfo
  {pages} {1049--1051} (\bibinfo {year} {1967})}\BibitemShut {NoStop}%
\bibitem [{\citenamefont {Mahan}(1967)}]{Mahan:1967dc}%
  \BibitemOpen
  \bibfield  {author} {\bibinfo {author} {\bibfnamefont {G}~\bibnamefont
  {Mahan}},\ }\bibfield  {title} {\enquote {\bibinfo {title} {{\sl Excitons in
  Metals: Infinite Hole Mass}},}\ }\href {\doibase 10.1103/PhysRev.163.612}
  {\bibfield  {journal} {\bibinfo  {journal} {Physical Review}\ }\textbf
  {\bibinfo {volume} {163}},\ \bibinfo {pages} {612--617} (\bibinfo {year}
  {1967})}\BibitemShut {NoStop}%
\bibitem [{\citenamefont {Nozi\`eres}\ and\ \citenamefont
  {de~Dominicis}(1969)}]{xray2}%
  \BibitemOpen
  \bibfield  {author} {\bibinfo {author} {\bibfnamefont {P.}~\bibnamefont
  {Nozi\`eres}}\ and\ \bibinfo {author} {\bibfnamefont {C.~T.}\ \bibnamefont
  {de~Dominicis}},\ }\bibfield  {title} {\enquote {\bibinfo {title} {{\sl
  Singularities in the X-Ray Absorption and Emission of Metals. III. One-Body
  Theory Exact Solution}},}\ }\href {\doibase 10.1103/PhysRev.178.1097}
  {\bibfield  {journal} {\bibinfo  {journal} {Phys. Rev.}\ }\textbf {\bibinfo
  {volume} {178}},\ \bibinfo {pages} {1097--1107} (\bibinfo {year}
  {1969})}\BibitemShut {NoStop}%
\bibitem [{\citenamefont {Anderson}(1995)}]{andersonirc}%
  \BibitemOpen
  \bibfield  {author} {\bibinfo {author} {\bibfnamefont {P.W.}\ \bibnamefont
  {Anderson}},\ }\bibfield  {title} {\enquote {\bibinfo {title} {{\sl
  “Infrared Catastrophe:” When Does It Trash Fermi Liquid Theory?}}}\ }in\
  \href@noop {} {\emph {\bibinfo {booktitle} {The Hubbard Model}}},\ \bibinfo
  {series} {NATO ASI Series}, Vol.\ \bibinfo {volume} {343},\ \bibinfo {editor}
  {edited by\ \bibinfo {editor} {\bibfnamefont {Dionys}\ \bibnamefont
  {Baeriswyl}}, \bibinfo {editor} {\bibfnamefont {DavidK.}\ \bibnamefont
  {Campbell}}, \bibinfo {editor} {\bibfnamefont {JoseM.P.}\ \bibnamefont
  {Carmelo}}, \bibinfo {editor} {\bibfnamefont {Francisco}\ \bibnamefont
  {Guinea}}, \ and\ \bibinfo {editor} {\bibfnamefont {Enrique}\ \bibnamefont
  {Louis}}}\ (\bibinfo  {publisher} {Springer US},\ \bibinfo {year} {1995})\
  pp.\ \bibinfo {pages} {217--225}\BibitemShut {NoStop}%
\bibitem [{\citenamefont {Imambekov}\ and\ \citenamefont
  {Glazman}(2009)}]{Imambekov:2009gi}%
  \BibitemOpen
  \bibfield  {author} {\bibinfo {author} {\bibfnamefont {Adilet}\ \bibnamefont
  {Imambekov}}\ and\ \bibinfo {author} {\bibfnamefont {Leonid~I}\ \bibnamefont
  {Glazman}},\ }\bibfield  {title} {\enquote {\bibinfo {title} {{Universal
  Theory of Nonlinear Luttinger Liquids}},}\ }\href
  {http://science.sciencemag.org/content/323/5911/228} {\bibfield  {journal}
  {\bibinfo  {journal} {Science (New York, NY)}\ }\textbf {\bibinfo {volume}
  {323}},\ \bibinfo {pages} {228--231} (\bibinfo {year} {2009})}\BibitemShut
  {NoStop}%
\bibitem [{\citenamefont {Fiete}(2009)}]{Fiete:2009gp}%
  \BibitemOpen
  \bibfield  {author} {\bibinfo {author} {\bibfnamefont {Gregory~A}\
  \bibnamefont {Fiete}},\ }\bibfield  {title} {\enquote {\bibinfo {title}
  {{\sl Singular responses of spin-incoherent Luttinger liquids}},}\ }\href
  {http://iopscience.iop.org/article/10.1088/0953-8984/21/19/193201} {\bibfield
   {journal} {\bibinfo  {journal} {J. Phys. Condens. Matter}\ }\textbf
  {\bibinfo {volume} {21}},\ \bibinfo {pages} {193201} (\bibinfo {year}
  {2009})}\BibitemShut {NoStop}%
\bibitem [{\citenamefont {Anderson}\ and\ \citenamefont {Yuval}(1969)}]{yuval}%
  \BibitemOpen
  \bibfield  {author} {\bibinfo {author} {\bibfnamefont {P.~W.}\ \bibnamefont
  {Anderson}}\ and\ \bibinfo {author} {\bibfnamefont {G.}~\bibnamefont
  {Yuval}},\ }\bibfield  {title} {\enquote {\bibinfo {title} {{\sl Exact
  Results in the Kondo Problem: Equivalence to a Classical One-Dimensional
  Coulomb Gas}},}\ }\href {\doibase 10.1103/PhysRevLett.23.89} {\bibfield
  {journal} {\bibinfo  {journal} {Phys. Rev. Lett.}\ }\textbf {\bibinfo
  {volume} {45}},\ \bibinfo {pages} {370} (\bibinfo {year} {1969})}\BibitemShut
  {NoStop}%
\bibitem [{\citenamefont {Anderson}\ and\ \citenamefont
  {Yuval}(1970)}]{yuval2}%
  \BibitemOpen
  \bibfield  {author} {\bibinfo {author} {\bibfnamefont {P.~W.}\ \bibnamefont
  {Anderson}}\ and\ \bibinfo {author} {\bibfnamefont {G.}~\bibnamefont
  {Yuval}},\ }\bibfield  {title} {\enquote {\bibinfo {title} {{\sl Exact
  Results for the Kondo Problem: One-Body Theory and Extension to Finite
  Temperature}},}\ }\href {\doibase 10.1103/PhysRevB.1.1522} {\bibfield
  {journal} {\bibinfo  {journal} {Phys. Rev. B}\ }\textbf {\bibinfo {volume}
  {1}},\ \bibinfo {pages} {1522} (\bibinfo {year} {1970})}\BibitemShut
  {NoStop}%
\bibitem [{\citenamefont {Anderson}\ \emph {et~al.}(1970)\citenamefont
  {Anderson}, \citenamefont {Yuval},\ and\ \citenamefont
  {Hamann}}]{Anderson:1970bv}%
  \BibitemOpen
  \bibfield  {author} {\bibinfo {author} {\bibfnamefont {P~W}\ \bibnamefont
  {Anderson}}, \bibinfo {author} {\bibfnamefont {G}~\bibnamefont {Yuval}}, \
  and\ \bibinfo {author} {\bibfnamefont {D}~\bibnamefont {Hamann}},\ }\bibfield
   {title} {\enquote {\bibinfo {title} {{\sl Exact Results in the Kondo
  Problem. II. Scaling Theory, Qualitatively Correct Solution, and Some New
  Results on One-Dimensional Classical Statistical Models}},}\ }\href
  {http://link.aps.org/doi/10.1103/PhysRevB.1.4464} {\bibfield  {journal}
  {\bibinfo  {journal} {Physical Review B}\ }\textbf {\bibinfo {volume} {1}},\
  \bibinfo {pages} {4464--4473} (\bibinfo {year} {1970})}\BibitemShut {NoStop}%
\bibitem [{\citenamefont {Anderson}(1968-69{\natexlab{b}})}]{kondoI}%
  \BibitemOpen
  \bibfield  {author} {\bibinfo {author} {\bibfnamefont {P.~W.}\ \bibnamefont
  {Anderson}},\ }\bibfield  {title} {\enquote {\bibinfo {title} {{\sl The Kondo
  Effect. I}},}\ }\href@noop {} {\bibfield  {journal} {\bibinfo  {journal}
  {Comments on Solid State Physics}\ }\textbf {\bibinfo {volume} {I}},\
  \bibinfo {pages} {31} (\bibinfo {year} {1968-69}{\natexlab{b}})}\BibitemShut
  {NoStop}%
\bibitem [{\citenamefont {Anderson}(1971)}]{kondoIII}%
  \BibitemOpen
  \bibfield  {author} {\bibinfo {author} {\bibfnamefont {P.~W.}\ \bibnamefont
  {Anderson}},\ }\bibfield  {title} {\enquote {\bibinfo {title} {{\sl The Kondo
  Effect III: The Wilderness-Mainly Theoretical}},}\ }\href@noop {} {\bibfield
  {journal} {\bibinfo  {journal} {Comments on Solid State Physics}\ }\textbf
  {\bibinfo {volume} {3}},\ \bibinfo {pages} {153} (\bibinfo {year}
  {1971})}\BibitemShut {NoStop}%
\bibitem [{\citenamefont {Anderson}(1973)}]{kondoIV}%
  \BibitemOpen
  \bibfield  {author} {\bibinfo {author} {\bibfnamefont {P.~W.}\ \bibnamefont
  {Anderson}},\ }\bibfield  {title} {\enquote {\bibinfo {title} {{\sl Kondo
  Effect IV:Out of the Wilderness}},}\ }\href@noop {} {\bibfield  {journal}
  {\bibinfo  {journal} {Comments on Solid State Physics}\ }\textbf {\bibinfo
  {volume} {5}},\ \bibinfo {pages} {73} (\bibinfo {year} {1973})}\BibitemShut
  {NoStop}%
\bibitem [{\citenamefont {Wilson}(1975)}]{revwilson}%
  \BibitemOpen
  \bibfield  {author} {\bibinfo {author} {\bibfnamefont {K.~G.}\ \bibnamefont
  {Wilson}},\ }\bibfield  {title} {\enquote {\bibinfo {title} {{\sl The
  renormalization group: Critical phenomena and the Kondo problem}},}\ }\href
  {\doibase 10.1103/RevModPhys.47.773} {\bibfield  {journal} {\bibinfo
  {journal} {Rev. Mod. Phys.}\ }\textbf {\bibinfo {volume} {47}},\ \bibinfo
  {pages} {773} (\bibinfo {year} {1975})}\BibitemShut {NoStop}%
\bibitem [{\citenamefont {Werner}\ \emph {et~al.}(2006)\citenamefont {Werner},
  \citenamefont {Comanac}, \citenamefont {de' Medici}, \citenamefont {Troyer},\
  and\ \citenamefont {Millis}}]{millisref}%
  \BibitemOpen
  \bibfield  {author} {\bibinfo {author} {\bibfnamefont {Philipp}\ \bibnamefont
  {Werner}}, \bibinfo {author} {\bibfnamefont {Armin}\ \bibnamefont {Comanac}},
  \bibinfo {author} {\bibfnamefont {Luca}\ \bibnamefont {de' Medici}}, \bibinfo
  {author} {\bibfnamefont {Matthias}\ \bibnamefont {Troyer}}, \ and\ \bibinfo
  {author} {\bibfnamefont {Andrew~J.}\ \bibnamefont {Millis}},\ }\bibfield
  {title} {\enquote {\bibinfo {title} {Continuous-time solver for quantum
  impurity models},}\ }\href {\doibase 10.1103/PhysRevLett.97.076405}
  {\bibfield  {journal} {\bibinfo  {journal} {Phys. Rev. Lett.}\ }\textbf
  {\bibinfo {volume} {97}},\ \bibinfo {pages} {076405} (\bibinfo {year}
  {2006})}\BibitemShut {NoStop}%
\bibitem [{\citenamefont {Nozi{\`e}res}(1974)}]{noziereskfl}%
  \BibitemOpen
  \bibfield  {author} {\bibinfo {author} {\bibfnamefont {P.}~\bibnamefont
  {Nozi{\`e}res}},\ }\bibfield  {title} {\enquote {\bibinfo {title} {{\sl A
  ``Fermi Liquid" Description of the Kondo Problem at Low Tempertures}},}\
  }\href {\doibase 10.1007/BF00654541} {\bibfield  {journal} {\bibinfo
  {journal} {Journal of Low Temperature Physics}\ }\textbf {\bibinfo {volume}
  {17}},\ \bibinfo {pages} {31--42} (\bibinfo {year} {1974})}\BibitemShut
  {NoStop}%
\bibitem [{\citenamefont {Kosterlitz}\ and\ \citenamefont
  {Thouless}(1973)}]{Kosterlitz:1973fc}%
  \BibitemOpen
  \bibfield  {author} {\bibinfo {author} {\bibfnamefont {J~M}\ \bibnamefont
  {Kosterlitz}}\ and\ \bibinfo {author} {\bibfnamefont {D~J}\ \bibnamefont
  {Thouless}},\ }\bibfield  {title} {\enquote {\bibinfo {title} {{\sl Ordering,
  metastability and phase transitions in two-dimensional systems}},}\ }\href
  {http://iopscience.iop.org/article/10.1088/0022-3719/6/7/010} {\bibfield
  {journal} {\bibinfo  {journal} {Journal of Physics C: Solid State Physics}\
  }\textbf {\bibinfo {volume} {6}},\ \bibinfo {pages} {1181} (\bibinfo {year}
  {1973})}\BibitemShut {NoStop}%
\bibitem [{\citenamefont {Andrei}(1980)}]{PhysRevLett.45.379}%
  \BibitemOpen
  \bibfield  {author} {\bibinfo {author} {\bibfnamefont {N.}~\bibnamefont
  {Andrei}},\ }\bibfield  {title} {\enquote {\bibinfo {title} {{\sl
  Diagonalization of the Kondo Hamiltonian}},}\ }\href
  {http://link.aps.org/doi/10.1103/PhysRevLett.45.379} {\bibfield  {journal}
  {\bibinfo  {journal} {Phys. Rev. Lett.}\ }\textbf {\bibinfo {volume} {45}},\
  \bibinfo {pages} {379--382} (\bibinfo {year} {1980})}\BibitemShut {NoStop}%
\bibitem [{\citenamefont {Weigman}(1980)}]{Vigman:1980uc}%
  \BibitemOpen
  \bibfield  {author} {\bibinfo {author} {\bibfnamefont {P~B}\ \bibnamefont
  {Weigman}},\ }\bibfield  {title} {\enquote {\bibinfo {title} {{\sl Exact
  solution of sd exchange model at T= 0}},}\ }\href
  {http://www.jetpletters.ac.ru/ps/1353/article_20434.shtml} {\bibfield
  {journal} {\bibinfo  {journal} {JETP Lett}\ }\textbf {\bibinfo {volume}
  {31}},\ \bibinfo {pages} {364--370} (\bibinfo {year} {1980})}\BibitemShut
  {NoStop}%
\bibitem [{\citenamefont {Haule}(2007)}]{hauleetalref}%
  \BibitemOpen
  \bibfield  {author} {\bibinfo {author} {\bibfnamefont {Kristjan}\
  \bibnamefont {Haule}},\ }\bibfield  {title} {\enquote {\bibinfo {title}
  {{\sl Quantum Monte Carlo impurity solver for cluster dynamical mean-field theory
  and electronic structure calculations with adjustable cluster base}},}\
  }\href {http://link.aps.org/doi/10.1103/PhysRevB.75.155113} {\bibfield
  {journal} {\bibinfo  {journal} {Phys. Rev. B}\ }\textbf {\bibinfo {volume}
  {75}},\ \bibinfo {pages} {155113} (\bibinfo {year} {2007})}\BibitemShut
  {NoStop}%
\bibitem [{\citenamefont {Anderson}(1970)}]{Anderson1970sg}%
  \BibitemOpen
  \bibfield  {author} {\bibinfo {author} {\bibfnamefont {P.~W.}\ \bibnamefont
  {Anderson}},\ }\bibfield  {title} {\enquote {\bibinfo {title} {{\sl
  Localisation theory and the Cu–Mn problem: Spin glasses. }},}\ }\href@noop
  {} {\bibfield  {journal} {\bibinfo  {journal} {Mater. Res. Bull.}\ }\textbf
  {\bibinfo {volume} {5}},\ \bibinfo {pages} {549} (\bibinfo {year}
  {1970})}\BibitemShut {NoStop}%
\bibitem [{\citenamefont {Edwards}\ and\ \citenamefont
  {Anderson}(1975)}]{edwardsanderson}%
  \BibitemOpen
  \bibfield  {author} {\bibinfo {author} {\bibfnamefont {S.~F.}\ \bibnamefont
  {Edwards}}\ and\ \bibinfo {author} {\bibfnamefont {P.~W.}\ \bibnamefont
  {Anderson}},\ }\bibfield  {title} {\enquote {\bibinfo {title} {{\sl Theory of
  spin glasses}},}\ }\href
  {http://iopscience.iop.org/article/10.1088/0305-4608/5/5/017/meta} {\bibfield
   {journal} {\bibinfo  {journal} {J. Phys F: Metal Phys}\ }\textbf {\bibinfo
  {volume} {5}},\ \bibinfo {pages} {965} (\bibinfo {year} {1975})}\BibitemShut
  {NoStop}%
\bibitem [{\citenamefont {Cannella}\ and\ \citenamefont
  {Mydosh}(1972)}]{Cannella:1972ea}%
  \BibitemOpen
  \bibfield  {author} {\bibinfo {author} {\bibfnamefont {V}~\bibnamefont
  {Cannella}}\ and\ \bibinfo {author} {\bibfnamefont {J~A}\ \bibnamefont
  {Mydosh}},\ }\bibfield  {title} {\enquote {\bibinfo {title} {{\sl Magnetic
  Ordering in Gold-Iron Alloys}},}\ }\href {\doibase 10.1103/PhysRevB.6.4220}
  {\bibfield  {journal} {\bibinfo  {journal} {Physical Review B}\ }\textbf
  {\bibinfo {volume} {6}},\ \bibinfo {pages} {4220--4237} (\bibinfo {year}
  {1972})}\BibitemShut {NoStop}%
\bibitem [{\citenamefont {Menth}\ \emph {et~al.}(1969)\citenamefont {Menth},
  \citenamefont {Buehler},\ and\ \citenamefont {Geballe}}]{smb6}%
  \BibitemOpen
  \bibfield  {author} {\bibinfo {author} {\bibfnamefont {A.}~\bibnamefont
  {Menth}}, \bibinfo {author} {\bibfnamefont {E.}~\bibnamefont {Buehler}}, \
  and\ \bibinfo {author} {\bibfnamefont {T.~H.}\ \bibnamefont {Geballe}},\
  }\bibfield  {title} {\enquote {\bibinfo {title} {{\sl Magnetic and
  Semiconducting Properties of $SmB_6$}},}\ }\href
  {http://link.aps.org/doi/10.1103/PhysRevLett.22.295} {\bibfield  {journal}
  {\bibinfo  {journal} {Phys. Rev. Lett}\ }\textbf {\bibinfo {volume} {22}},\
  \bibinfo {pages} {295} (\bibinfo {year} {1969})}\BibitemShut {NoStop}%
\bibitem [{\citenamefont {Andres}\ \emph {et~al.}(1975)\citenamefont {Andres},
  \citenamefont {Graebner},\ and\ \citenamefont {Ott}}]{ott75}%
  \BibitemOpen
  \bibfield  {author} {\bibinfo {author} {\bibfnamefont {K.}~\bibnamefont
  {Andres}}, \bibinfo {author} {\bibfnamefont {J.}~\bibnamefont {Graebner}}, \
  and\ \bibinfo {author} {\bibfnamefont {H.~R.}\ \bibnamefont {Ott}},\
  }\bibfield  {title} {\enquote {\bibinfo {title} {{\sl 4f-Virtual-Bound-State
  Formation in $CeAl_3$ at Low Temperatures}},}\ }\href
  {http://link.aps.org/doi/10.1103/PhysRevLett.35.1779} {\bibfield  {journal}
  {\bibinfo  {journal} {Phys. Rev. Lett.}\ }\textbf {\bibinfo {volume} {35}},\
  \bibinfo {pages} {1779} (\bibinfo {year} {1975})}\BibitemShut {NoStop}%
\bibitem [{\citenamefont {Steglich}\ \emph {et~al.}(1979)\citenamefont
  {Steglich}, \citenamefont {Aarts}, \citenamefont {Bredl}, \citenamefont
  {Leike}, \citenamefont {Franz},\ and\ \citenamefont {Sch{\"
  a}fer}}]{steglich}%
  \BibitemOpen
  \bibfield  {author} {\bibinfo {author} {\bibfnamefont {F.}~\bibnamefont
  {Steglich}}, \bibinfo {author} {\bibfnamefont {J.}~\bibnamefont {Aarts}},
  \bibinfo {author} {\bibfnamefont {C.~D.}\ \bibnamefont {Bredl}}, \bibinfo
  {author} {\bibfnamefont {W.}~\bibnamefont {Leike}}, \bibinfo {author}
  {\bibfnamefont {D.~E. Meshida~W.}\ \bibnamefont {Franz}}, \ and\ \bibinfo
  {author} {\bibfnamefont {H.}~\bibnamefont {Sch{\" a}fer}},\ }\bibfield
  {title} {\enquote {\bibinfo {title} {{\sl Superconductivity in the Presence
  of Strong Pauli Paramagnetism: $CeCu_ {2} Si_ {2}$}},}\ }\href
  {http://link.aps.org/doi/10.1103/PhysRevLett.43.1892} {\bibfield  {journal}
  {\bibinfo  {journal} {Phys. Rev. Lett}\ }\textbf {\bibinfo {volume} {43}},\
  \bibinfo {pages} {1892} (\bibinfo {year} {1979})}\BibitemShut {NoStop}%
\bibitem [{\citenamefont {Anderson}(1977)}]{epilogue}%
  \BibitemOpen
  \bibfield  {author} {\bibinfo {author} {\bibfnamefont {P.~W.}\ \bibnamefont
  {Anderson}},\ }\bibfield  {title} {\enquote {\bibinfo {title} {{\sl
  Epilogue}},}\ }in\ \href@noop {} {\emph {\bibinfo {booktitle} {Valence
  Instabilities and Narrow-B and Phenomena}}},\ \bibinfo {editor} {edited by\
  \bibinfo {editor} {\bibfnamefont {R.~D.}\ \bibnamefont {Parks}}}\ (\bibinfo
  {publisher} {Plenum, NY},\ \bibinfo {year} {1977})\ pp.\ \bibinfo {pages}
  {389--396}\BibitemShut {NoStop}%
\bibitem [{\citenamefont {Bucher}\ \emph {et~al.}(1975)\citenamefont {Bucher},
  \citenamefont {Maita}, \citenamefont {Hull}, \citenamefont {Fulton},\ and\
  \citenamefont {Cooper}}]{bucher}%
  \BibitemOpen
  \bibfield  {author} {\bibinfo {author} {\bibfnamefont {E.}~\bibnamefont
  {Bucher}}, \bibinfo {author} {\bibfnamefont {J.~P.}\ \bibnamefont {Maita}},
  \bibinfo {author} {\bibfnamefont {G.~W.}\ \bibnamefont {Hull}}, \bibinfo
  {author} {\bibfnamefont {R.~C.}\ \bibnamefont {Fulton}}, \ and\ \bibinfo
  {author} {\bibfnamefont {A.~S.}\ \bibnamefont {Cooper}},\ }\bibfield  {title}
  {\enquote {\bibinfo {title} {{\sl Electronic properties of beryllides of the
  rare earth and some actinides}},}\ }\href {\doibase 10.1103/PhysRevB.11.440}
  {\bibfield  {journal} {\bibinfo  {journal} {Phys. Rev. B}\ }\textbf {\bibinfo
  {volume} {11}},\ \bibinfo {pages} {440} (\bibinfo {year} {1975})}\BibitemShut
  {NoStop}%
\bibitem [{\citenamefont {Ramakrishnan}(1981)}]{ramalargeN}%
  \BibitemOpen
  \bibfield  {author} {\bibinfo {author} {\bibfnamefont {T.~V.}\ \bibnamefont
  {Ramakrishnan}},\ }\bibfield  {title} {\enquote {\bibinfo {title} {{}},}\
  }in\ \href@noop {} {\emph {\bibinfo {booktitle} {{\sl Valence Fluctuations in
  Solids}}}},\ \bibinfo {editor} {edited by\ \bibinfo {editor} {\bibnamefont
  {{L. M. Falicov and W. Hanke and M. P. Maple}}}}\ (\bibinfo  {publisher}
  {{North Holland}},\ \bibinfo {address} {Amsterdam},\ \bibinfo {year} {1981})\
  p.~\bibinfo {pages} {{13}}\BibitemShut {NoStop}%
\bibitem [{\citenamefont {Ramakrishnan}\ and\ \citenamefont
  {Sur}(1982)}]{ramakrishnansur}%
  \BibitemOpen
  \bibfield  {author} {\bibinfo {author} {\bibfnamefont {T.~V.}\ \bibnamefont
  {Ramakrishnan}}\ and\ \bibinfo {author} {\bibfnamefont {K.}~\bibnamefont
  {Sur}},\ }\bibfield  {title} {\enquote {\bibinfo {title} {Theory of a
  mixed-valent impurity},}\ }\href
  {http://link.aps.org/doi/10.1103/PhysRevB.26.1798} {\bibfield  {journal}
  {\bibinfo  {journal} {Phys. Rev. B}\ }\textbf {\bibinfo {volume} {26}},\
  \bibinfo {pages} {1798--1811} (\bibinfo {year} {1982})}\BibitemShut {NoStop}%
\bibitem [{\citenamefont {Gunnarsson}\ and\ \citenamefont {Sch{\"
  o}nhammer}(1983)}]{gunnarson}%
  \BibitemOpen
  \bibfield  {author} {\bibinfo {author} {\bibfnamefont {O.}~\bibnamefont
  {Gunnarsson}}\ and\ \bibinfo {author} {\bibfnamefont {K.}~\bibnamefont
  {Sch{\" o}nhammer}},\ }\bibfield  {title} {\enquote {\bibinfo {title} {{\sl
  Electron spectroscopies for Ce compounds in the impurity model}},}\ }\href
  {\doibase 10.1103/PhysRevB.28.4315} {\bibfield  {journal} {\bibinfo
  {journal} {Phys. Rev. B}\ }\textbf {\bibinfo {volume} {28}},\ \bibinfo
  {pages} {4315} (\bibinfo {year} {1983})}\BibitemShut {NoStop}%
\bibitem [{\citenamefont {Zhang}\ and\ \citenamefont {Lee}(1983)}]{zhanglee}%
  \BibitemOpen
  \bibfield  {author} {\bibinfo {author} {\bibfnamefont {F.~C.}\ \bibnamefont
  {Zhang}}\ and\ \bibinfo {author} {\bibfnamefont {T.~K.}\ \bibnamefont
  {Lee}},\ }\bibfield  {title} {\enquote {\bibinfo {title} {{\sl $\frac{1}{N}$
  expansion for the degenerate Anderson model in the mixed-valence regime}},}\
  }\href {\doibase 10.1103/PhysRevB.28.33} {\bibfield  {journal} {\bibinfo
  {journal} {Phys. Rev. B}\ }\textbf {\bibinfo {volume} {28}},\ \bibinfo
  {pages} {33--38} (\bibinfo {year} {1983})}\BibitemShut {NoStop}%
\bibitem [{\citenamefont {Read}\ and\ \citenamefont
  {Newns}(1983{\natexlab{a}})}]{read83_1}%
  \BibitemOpen
  \bibfield  {author} {\bibinfo {author} {\bibfnamefont {N.}~\bibnamefont
  {Read}}\ and\ \bibinfo {author} {\bibfnamefont {D.M.}\ \bibnamefont
  {Newns}},\ }\bibfield  {title} {\enquote {\bibinfo {title} {{\sl On the
  solution of the Coqblin-Schreiffer Hamiltonian by the large-N expansion
  technique}},}\ }\href
  {http://iopscience.iop.org/article/10.1088/0022-3719/16/17/014/meta}
  {\bibfield  {journal} {\bibinfo  {journal} {J. Phys. C}\ }\textbf {\bibinfo
  {volume} {16}},\ \bibinfo {pages} {3273--3295} (\bibinfo {year}
  {1983}{\natexlab{a}})}\BibitemShut {NoStop}%
\bibitem [{\citenamefont {Read}\ and\ \citenamefont
  {Newns}(1983{\natexlab{b}})}]{read83_v2}%
  \BibitemOpen
  \bibfield  {author} {\bibinfo {author} {\bibfnamefont {N.}~\bibnamefont
  {Read}}\ and\ \bibinfo {author} {\bibfnamefont {D.~M.}\ \bibnamefont
  {Newns}},\ }\bibfield  {title} {\enquote {\bibinfo {title} {{\sl A new
  functional integral formalism for the degenerate Anderson model}},}\ }\href
  {http://iopscience.iop.org/article/10.1088/0022-3719/16/29/007/meta}
  {\bibfield  {journal} {\bibinfo  {journal} {J. Phys. C}\ }\textbf {\bibinfo
  {volume} {29}},\ \bibinfo {pages} {L1055} (\bibinfo {year}
  {1983}{\natexlab{b}})}\BibitemShut {NoStop}%
\bibitem [{\citenamefont {Coleman}(1983)}]{me1983}%
  \BibitemOpen
  \bibfield  {author} {\bibinfo {author} {\bibfnamefont {P.}~\bibnamefont
  {Coleman}},\ }\bibfield  {title} {\enquote {\bibinfo {title} {{\sl 1/N
  expansion for the Kondo lattice}},}\ }\href
  {http://link.aps.org/doi/10.1103/PhysRevB.28.5255} {\bibfield  {journal}
  {\bibinfo  {journal} {Phys. Rev. B.}\ }\textbf {\bibinfo {volume} {28}},\
  \bibinfo {pages} {5255} (\bibinfo {year} {1983})}\BibitemShut {NoStop}%
\bibitem [{\citenamefont {Nakajima}\ \emph {et~al.}(2004)\citenamefont
  {Nakajima}, \citenamefont {Izawa}, \citenamefont {Matsuda}, \citenamefont
  {Uji}, \citenamefont {Terashima}, \citenamefont {Shishido}, \citenamefont
  {Settai}, \citenamefont {Onuki},\ and\ \citenamefont
  {Kontani}}]{Nakajima:2004wy}%
  \BibitemOpen
  \bibfield  {author} {\bibinfo {author} {\bibfnamefont {Y}~\bibnamefont
  {Nakajima}}, \bibinfo {author} {\bibfnamefont {K}~\bibnamefont {Izawa}},
  \bibinfo {author} {\bibfnamefont {Y}~\bibnamefont {Matsuda}}, \bibinfo
  {author} {\bibfnamefont {S}~\bibnamefont {Uji}}, \bibinfo {author}
  {\bibfnamefont {T}~\bibnamefont {Terashima}}, \bibinfo {author}
  {\bibfnamefont {H}~\bibnamefont {Shishido}}, \bibinfo {author} {\bibfnamefont
  {R}~\bibnamefont {Settai}}, \bibinfo {author} {\bibfnamefont {Y}~\bibnamefont
  {Onuki}}, \ and\ \bibinfo {author} {\bibfnamefont {H}~\bibnamefont
  {Kontani}},\ }\bibfield  {title} {\enquote {\bibinfo {title} {{\sl
  Normal-state Hall Angle and Magnetoresistance in quasi-2D Heavy Fermion
  CeCoIn$_5$ near a Quantum Critical Point}},}\ }\href
  {http://journals.jps.jp/doi/abs/10.1143/JPSJ.73.5} {\bibfield  {journal}
  {\bibinfo  {journal} {Journal Of The Physical Society Of Japan}\ }\textbf
  {\bibinfo {volume} {73}},\ \bibinfo {pages} {5} (\bibinfo {year}
  {2004})}\BibitemShut {NoStop}%
\bibitem [{\citenamefont {Osheroff}\ \emph {et~al.}(1972)\citenamefont
  {Osheroff}, \citenamefont {Richardson},\ and\ \citenamefont
  {Lee}}]{Osheroff1972}%
  \BibitemOpen
  \bibfield  {author} {\bibinfo {author} {\bibfnamefont {D.~D.}\ \bibnamefont
  {Osheroff}}, \bibinfo {author} {\bibfnamefont {R.~C.}\ \bibnamefont
  {Richardson}}, \ and\ \bibinfo {author} {\bibfnamefont {D.~M.}\ \bibnamefont
  {Lee}},\ }\bibfield  {title} {\enquote {\bibinfo {title} {{\sl "Evidence for
  a New Phase of Solid ${\mathrm{He}}^{3}$"}},}\ }\href
  {http://link.aps.org/doi/10.1103/PhysRevLett.28.885} {\bibfield  {journal}
  {\bibinfo  {journal} {Phys. Rev. Lett.}\ }\textbf {\bibinfo {volume} {28}},\
  \bibinfo {pages} {885--888} (\bibinfo {year} {1972})}\BibitemShut {NoStop}%
\bibitem [{\citenamefont {J\'erome}\ \emph {et~al.}(1980)\citenamefont
  {J\'erome}, \citenamefont {Mazaud},\ and\ \citenamefont
  {Bechgaard}}]{jerome}%
  \BibitemOpen
  \bibfield  {author} {\bibinfo {author} {\bibfnamefont {D.}~\bibnamefont
  {J\'erome}}, \bibinfo {author} {\bibfnamefont {M.}~\bibnamefont {Mazaud},
  \bibfnamefont {A.~Ribault}}, \ and\ \bibinfo {author} {\bibfnamefont
  {K.}~\bibnamefont {Bechgaard}},\ }\bibfield  {title} {\enquote {\bibinfo
  {title} {{\sl Superconductivity in a synthetic organic conductor:
  (TMTSF)$_2$PF$_6$}},}\ }\href {\doibase 10.1051/jphyslet:0198000410409500}
  {\bibfield  {journal} {\bibinfo  {journal} {J. Phys. Lett. (Paris)}\ }\textbf
  {\bibinfo {volume} {41}},\ \bibinfo {pages} {L95} (\bibinfo {year}
  {1980})}\BibitemShut {NoStop}%
\bibitem [{\citenamefont {Bednorz}\ and\ \citenamefont
  {Muller}(1986)}]{bednorz}%
  \BibitemOpen
  \bibfield  {author} {\bibinfo {author} {\bibfnamefont {J~G}\ \bibnamefont
  {Bednorz}}\ and\ \bibinfo {author} {\bibfnamefont {K~A}\ \bibnamefont
  {Muller}},\ }\bibfield  {title} {\enquote {\bibinfo {title} {{\sl Possible
  high Tc superconductivity in the Ba-La-Cu-O system}},}\ }\href
  {http://dx.doi.org/10.1007/BF01303701} {\bibfield  {journal} {\bibinfo
  {journal} {Z.Phys.}\ }\textbf {\bibinfo {volume} {B64}},\ \bibinfo {pages}
  {189--193} (\bibinfo {year} {1986})}\BibitemShut {NoStop}%
\bibitem [{\citenamefont {Kamihara}\ \emph {et~al.}(2006)\citenamefont
  {Kamihara}, \citenamefont {Hiramatsu}, \citenamefont {Hirano}, \citenamefont
  {Kawamura}, \citenamefont {Yanagi}, \citenamefont {Kamiya},\ and\
  \citenamefont {Hosono}}]{hosono}%
  \BibitemOpen
  \bibfield  {author} {\bibinfo {author} {\bibfnamefont {Yoichi}\ \bibnamefont
  {Kamihara}}, \bibinfo {author} {\bibfnamefont {Hidenori}\ \bibnamefont
  {Hiramatsu}}, \bibinfo {author} {\bibfnamefont {Masahiro}\ \bibnamefont
  {Hirano}}, \bibinfo {author} {\bibfnamefont {Ryuto}\ \bibnamefont
  {Kawamura}}, \bibinfo {author} {\bibfnamefont {Hiroshi}\ \bibnamefont
  {Yanagi}}, \bibinfo {author} {\bibfnamefont {Toshio}\ \bibnamefont {Kamiya}},
  \ and\ \bibinfo {author} {\bibfnamefont {Hideo}\ \bibnamefont {Hosono}},\
  }\bibfield  {title} {\enquote {\bibinfo {title} {{\sl Iron-Based Layered
  Superconductor: LaOFeP}},}\ }\href
  {http://pubs.acs.org/doi/abs/10.1021/ja063355c} {\bibfield  {journal}
  {\bibinfo  {journal} {Journal of the American Chemical Society}\ }\textbf
  {\bibinfo {volume} {128}},\ \bibinfo {pages} {10012--10013} (\bibinfo {year}
  {2006})}\BibitemShut {NoStop}%
\bibitem [{\citenamefont {Anderson}(1990)}]{philstrange1}%
  \BibitemOpen
  \bibfield  {author} {\bibinfo {author} {\bibfnamefont {P~W}\ \bibnamefont
  {Anderson}},\ }\bibfield  {title} {\enquote {\bibinfo {title}
  {{\sl Luttinger-liquid
  behavior of the normal metallic state of the 2D Hubbard model}},}\ }\href
  {\doibase 10.1103/PhysRevLett.64.1839} {\bibfield  {journal} {\bibinfo
  {journal} {Physical Review Letters}\ }\textbf {\bibinfo {volume} {64}},\
  \bibinfo {pages} {1839--1841} (\bibinfo {year} {1990})}\BibitemShut {NoStop}%
\bibitem [{\citenamefont {Anderson}(2006)}]{philstrange2}%
  \BibitemOpen
  \bibfield  {author} {\bibinfo {author} {\bibfnamefont {P~W}\ \bibnamefont
  {Anderson}},\ }\bibfield  {title} {\enquote {\bibinfo {title} {{\sl The
  'strange metal' is a projected Fermi liquid with edge singularities}},}\
  }\href {\doibase 10.1038/nphys388} {\bibfield  {journal} {\bibinfo  {journal}
  {Nature Physics}\ }\textbf {\bibinfo {volume} {2}},\ \bibinfo {pages}
  {626--630} (\bibinfo {year} {2006})}\BibitemShut {NoStop}%
\bibitem [{\citenamefont {Anderson}(2008)}]{phil_hidden}%
  \BibitemOpen
  \bibfield  {author} {\bibinfo {author} {\bibfnamefont {P.~W.}\ \bibnamefont
  {Anderson}},\ }\bibfield  {title} {\enquote {\bibinfo {title} {{\sl Hidden
  Fermi liquid: The secret of high-T$_c$ cuprates}},}\ }\href
  {http://link.aps.org/doi/10.1103/PhysRevB.78.174505} {\bibfield  {journal}
  {\bibinfo  {journal} {Phys. Rev. B}\ }\textbf {\bibinfo {volume} {78}},\
  \bibinfo {pages} {174505} (\bibinfo {year} {2008})}\BibitemShut {NoStop}%
\bibitem [{\citenamefont {Anderson}(2010)}]{phil_hidden2}%
  \BibitemOpen
  \bibfield  {author} {\bibinfo {author} {\bibfnamefont {Philip~W}\
  \bibnamefont {Anderson}},\ }\bibfield  {title} {\enquote {\bibinfo {title}
  {{\sl Fermi Sea of Heavy Electrons (a Kondo Lattice) is Never a Fermi Liquid}},}\
  }\href {\doibase 10.1103/PhysRevLett.104.176403} {\bibfield  {journal}
  {\bibinfo  {journal} {Physical Review Letters}\ }\textbf {\bibinfo {volume}
  {104}},\ \bibinfo {pages} {176403} (\bibinfo {year} {2010})}\BibitemShut
  {NoStop}%
\bibitem [{\citenamefont {Chien}\ \emph {et~al.}(1991)\citenamefont {Chien},
  \citenamefont {Wang},\ and\ \citenamefont {Ong}}]{ong}%
  \BibitemOpen
  \bibfield  {author} {\bibinfo {author} {\bibfnamefont {T~R}\ \bibnamefont
  {Chien}}, \bibinfo {author} {\bibfnamefont {Z~Z}\ \bibnamefont {Wang}}, \
  and\ \bibinfo {author} {\bibfnamefont {N~P}\ \bibnamefont {Ong}},\ }\bibfield
   {title} {\enquote {\bibinfo {title} {{\sl Effect of Zn impurities on the
  normal-state Hall angle in single-crystal
  ${\mathrm{YBa}}_{2}$${\mathrm{Cu}}_{3\mathrm{-}\mathit{x}}$${\mathrm{Zn}}_{\mathit{x}}$
  ${\mathrm{O}}_{7\mathrm{-}\delta}$}},}\ }\href
  {http://link.aps.org/doi/10.1103/PhysRevLett.67.2088} {\bibfield  {journal}
  {\bibinfo  {journal} {Physical Review Letters}\ }\textbf {\bibinfo {volume}
  {67}},\ \bibinfo {pages} {2088} (\bibinfo {year} {1991})}\BibitemShut
  {NoStop}%
\bibitem [{\citenamefont {Pfleiderer}\ \emph {et~al.}(2001)\citenamefont
  {Pfleiderer}, \citenamefont {Julian},\ and\ \citenamefont
  {Lonzarich}}]{lonzarichmnsi}%
  \BibitemOpen
  \bibfield  {author} {\bibinfo {author} {\bibfnamefont {C}~\bibnamefont
  {Pfleiderer}}, \bibinfo {author} {\bibfnamefont {S~R}\ \bibnamefont
  {Julian}}, \ and\ \bibinfo {author} {\bibfnamefont {G~G}\ \bibnamefont
  {Lonzarich}},\ }\bibfield  {title} {\enquote {\bibinfo {title} {{\sl
  Non-Fermi-liquid nature of the normal state of itinerant-electron
  ferromagnets}},}\ }\href {\doibase 10.1038/35106527} {\bibfield  {journal}
  {\bibinfo  {journal} {Nature}\ }\textbf {\bibinfo {volume} {414}},\ \bibinfo
  {pages} {427--430} (\bibinfo {year} {2001})}\BibitemShut {NoStop}%
\bibitem [{\citenamefont {Tanatar}\ \emph {et~al.}(2007)\citenamefont
  {Tanatar}, \citenamefont {Paglione}, \citenamefont {Petrovic},\ and\
  \citenamefont {Taillefer}}]{Tanatar}%
  \BibitemOpen
  \bibfield  {author} {\bibinfo {author} {\bibfnamefont {Makariy~A}\
  \bibnamefont {Tanatar}}, \bibinfo {author} {\bibfnamefont {Johnpierre}\
  \bibnamefont {Paglione}}, \bibinfo {author} {\bibfnamefont {Cedomir}\
  \bibnamefont {Petrovic}}, \ and\ \bibinfo {author} {\bibfnamefont {Louis}\
  \bibnamefont {Taillefer}},\ }\bibfield  {title} {\enquote {\bibinfo {title}
  {{\sl Anisotropic Violation of the Wiedemann-Franz Law at a Quantum Critical
  Point}},}\ }\href {\doibase 10.1126/science.1140762} {\bibfield  {journal}
  {\bibinfo  {journal} {Science (New York, NY)}\ }\textbf {\bibinfo {volume}
  {316}},\ \bibinfo {pages} {1320--1322} (\bibinfo {year} {2007})}\BibitemShut
  {NoStop}%
\bibitem [{\citenamefont {Matsumoto}\ \emph {et~al.}(2011)\citenamefont
  {Matsumoto}, \citenamefont {Nakatsuji}, \citenamefont {Kuga}, \citenamefont
  {Karaki}, \citenamefont {Horie}, \citenamefont {Shimura}, \citenamefont
  {Sakakibara}, \citenamefont {Nevidomskyy},\ and\ \citenamefont
  {Coleman}}]{satoru}%
  \BibitemOpen
  \bibfield  {author} {\bibinfo {author} {\bibfnamefont {Yosuke}\ \bibnamefont
  {Matsumoto}}, \bibinfo {author} {\bibfnamefont {Satoru}\ \bibnamefont
  {Nakatsuji}}, \bibinfo {author} {\bibfnamefont {Kentaro}\ \bibnamefont
  {Kuga}}, \bibinfo {author} {\bibfnamefont {Yoshitomo}\ \bibnamefont
  {Karaki}}, \bibinfo {author} {\bibfnamefont {Naoki}\ \bibnamefont {Horie}},
  \bibinfo {author} {\bibfnamefont {Yasuyuki}\ \bibnamefont {Shimura}},
  \bibinfo {author} {\bibfnamefont {Toshiro}\ \bibnamefont {Sakakibara}},
  \bibinfo {author} {\bibfnamefont {Andriy~H.}\ \bibnamefont {Nevidomskyy}}, \
  and\ \bibinfo {author} {\bibfnamefont {Piers}\ \bibnamefont {Coleman}},\
  }\bibfield  {title} {\enquote {\bibinfo {title} {{\sl Quantum Criticality
  Without Tuning in the Mixed Valence Compound $\beta$-YbAlB4}},}\ }\href
  {\doibase 10.1126/science.1197531} {\bibfield  {journal} {\bibinfo  {journal}
  {Science (New York, NY)}\ }\textbf {\bibinfo {volume} {331}},\ \bibinfo
  {pages} {316--319} (\bibinfo {year} {2011})}\BibitemShut {NoStop}%
\bibitem [{\citenamefont {PW}(2007)}]{PW:2007ei}%
  \BibitemOpen
  \bibfield  {author} {\bibinfo {author} {\bibfnamefont {Anderson}\
  \bibnamefont {PW}},\ }\bibfield  {title} {\enquote {\bibinfo {title} {{\sl
  Physics. Is there glue in cuprate superconductors?}}}\ }\href
  {http://science.sciencemag.org/content/316/5832/1705.full} {\bibfield
  {journal} {\bibinfo  {journal} {Science (New York, NY)}\ }\textbf {\bibinfo
  {volume} {316}},\ \bibinfo {pages} {1705} (\bibinfo {year}
  {2007})}\BibitemShut {NoStop}%
\bibitem [{\citenamefont {Anderson}(1991)}]{Anderson:1991wz}%
  \BibitemOpen
  \bibfield  {author} {\bibinfo {author} {\bibfnamefont {P.~W.}\ \bibnamefont
  {Anderson}},\ }\bibfield  {title} {\enquote {\bibinfo {title} {{\sl Hall
  effect in the two-dimensional Luttinger liquid}},}\ }\href
  {http://link.aps.org/doi/10.1103/PhysRevLett.67.2092} {\bibfield  {journal}
  {\bibinfo  {journal} {Phys. Rev. Lett.}\ }\textbf {\bibinfo {volume} {67}},\
  \bibinfo {pages} {2092--2094} (\bibinfo {year} {1991})}\BibitemShut {NoStop}%
\bibitem [{\citenamefont {Mikheev}\ \emph {et~al.}(2015)\citenamefont
  {Mikheev}, \citenamefont {Freeze}, \citenamefont {Isaac}, \citenamefont
  {Cain},\ and\ \citenamefont {Stemmer}}]{Mikheev:2015cb}%
  \BibitemOpen
  \bibfield  {author} {\bibinfo {author} {\bibfnamefont {Evgeny}\ \bibnamefont
  {Mikheev}}, \bibinfo {author} {\bibfnamefont {Christopher~R}\ \bibnamefont
  {Freeze}}, \bibinfo {author} {\bibfnamefont {Brandon~J}\ \bibnamefont
  {Isaac}}, \bibinfo {author} {\bibfnamefont {Tyler~A}\ \bibnamefont {Cain}}, \
  and\ \bibinfo {author} {\bibfnamefont {Susanne}\ \bibnamefont {Stemmer}},\
  }\bibfield  {title} {\enquote {\bibinfo {title} {{\sl Separation of transport
  lifetimes in SrTiO3-based two-dimensional electron liquids}},}\ }\href
  {\doibase 10.1103/PhysRevB.91.165125} {\bibfield  {journal} {\bibinfo
  {journal} {Phys. Rev. B}\ }\textbf {\bibinfo {volume} {91}},\ \bibinfo
  {pages} {165125} (\bibinfo {year} {2015})}\BibitemShut {NoStop}%
\bibitem [{\citenamefont {Coleman}\ \emph {et~al.}(1996)\citenamefont
  {Coleman}, \citenamefont {Schofield},\ and\ \citenamefont
  {Tsvelik}}]{twotimes}%
  \BibitemOpen
  \bibfield  {author} {\bibinfo {author} {\bibfnamefont {Piers}\ \bibnamefont
  {Coleman}}, \bibinfo {author} {\bibfnamefont {A.~J.}\ \bibnamefont
  {Schofield}}, \ and\ \bibinfo {author} {\bibfnamefont {A.~M.}\ \bibnamefont
  {Tsvelik}},\ }\bibfield  {title} {\enquote {\bibinfo {title} {{\sl
  Phenomenological transport equation for the cuprate metals}},}\ }\href
  {\doibase 10.1103/PhysRevLett.76.1324} {\bibfield  {journal} {\bibinfo
  {journal} {Physical Review Letters}\ }\textbf {\bibinfo {volume} {76}},\
  \bibinfo {pages} {1324} (\bibinfo {year} {1996})}\BibitemShut {NoStop}%
\bibitem [{\citenamefont {Hussey}(2008)}]{Hussey:2008tw}%
  \BibitemOpen
  \bibfield  {author} {\bibinfo {author} {\bibfnamefont {N~E}\ \bibnamefont
  {Hussey}},\ }\bibfield  {title} {\enquote {\bibinfo {title} {\sl Phenomenology of
  the normal state in-plane transport properties of high- T$_{c}$ cuprates},}\
  }\href {http://stacks.iop.org/0953-8984/20/i=12/a=123201} {\bibfield
  {journal} {\bibinfo  {journal} {Journal of Physics: Condensed Matter}\
  }\textbf {\bibinfo {volume} {20}},\ \bibinfo {pages} {123201} (\bibinfo
  {year} {2008})}\BibitemShut {NoStop}%
\bibitem [{\citenamefont {Blake}\ and\ \citenamefont
  {Donos}(2015)}]{Blake:2015gc}%
  \BibitemOpen
  \bibfield  {author} {\bibinfo {author} {\bibfnamefont {Mike}\ \bibnamefont
  {Blake}}\ and\ \bibinfo {author} {\bibfnamefont {Aristomenis}\ \bibnamefont
  {Donos}},\ }\bibfield  {title} {\enquote {\bibinfo {title} {{\sl Quantum
  Critical Transport and the Hall Angle in Holographic Models}},}\ }\href
  {\doibase 10.1103/PhysRevLett.114.021601} {\bibfield  {journal} {\bibinfo
  {journal} {Phys. Rev. Lett.}\ }\textbf {\bibinfo {volume} {114}},\ \bibinfo
  {pages} {021601} (\bibinfo {year} {2015})}\BibitemShut {NoStop}%
\end{thebibliography}
